\documentclass[sigplan,10pt,screen]{acmart}

\definecolor{blush}{rgb}{0.87, 0.36, 0.51}
\definecolor{bondiblue}{rgb}{0.0, 0.58, 0.71}

\usepackage{pifont}
\usepackage{algorithm}
\usepackage[noend]{algpseudocode}
\usepackage{threeparttable}
\usepackage{tablefootnote}
\usepackage{adjustbox}
\usepackage{makecell}
\usepackage{amsthm}
\usepackage{enumitem}
\usepackage{subcaption}
\renewcommand\footnotetextcopyrightpermission[1]{}
\AtBeginDocument{%
  \providecommand\BibTeX{{%
    \normalfont B\kern-0.5em{\scshape i\kern-0.25em b}\kern-0.8em\TeX}}}

\setcopyright{acmcopyright}
\copyrightyear{2018}
\acmYear{2018}
\acmDOI{XXXXXXX.XXXXXXX}

\acmConference[Conference acronym 'XX]{Make sure to enter the correct
  conference title from your rights confirmation email}{June 03--05,
  2018}{Woodstock, NY}
\acmPrice{15.00}
\acmISBN{978-1-4503-XXXX-X/18/06}

\begin{document}
\begin{sloppypar}

\setlength{\abovecaptionskip}{5pt}
\setlength{\intextsep}{7pt plus 2pt minus 2pt}
\setlength{\textfloatsep}{7pt plus 2pt minus 2pt}

\title{Matryoshka: Optimization of Dynamic Diverse Quantum Chemistry Systems via Elastic Parallelism Transformation}

\renewcommand{\shorttitle}{Matryoshka}


\author{Tuowei Wang}\authornote{Work done during an internship at Microsoft Research.}
\affiliation{%
  \institution{Microsoft Research} 
  \institution{Tsinghua University}
  \country{Beijing, China}
}
\author{Kun Li}\authornote{Corresponding author.}
\affiliation{%
  \institution{Microsoft Research} 
  \country{Beijing, China}
}
\author{Donglin Bai}
\affiliation{%
  \institution{Microsoft Research} 
  \country{Beijing, China}
}
\author{Fusong Ju}
\affiliation{%
  \institution{Microsoft Research} 
  \country{Beijing, China}
}
\author{Leo Xia}
\affiliation{%
  \institution{Microsoft Research} 
  \country{Beijing, China}
}
\author{Ju Ren}
\affiliation{%
  \institution{Tsinghua University}
  \country{Beijing, China}
}
\author{Yaoxue Zhang}
\affiliation{%
  \institution{Tsinghua University}
  \country{Beijing, China}
}
\author{Ting Cao}
\affiliation{%
  \institution{Microsoft Research}
  \country{Beijing, China}
}
\author{Mao Yang}
\affiliation{%
  \institution{Microsoft Research}
  \country{Beijing, China}
}

\renewcommand{\shortauthors}{Tuo and Kun, et al.}

\begin{abstract}
  AI infrastructures, predominantly GPUs, have delivered remarkable performance gains for deep learning. Conversely, scientific computing, exemplified by quantum chemistry systems, suffers from dynamic diversity, where computational patterns are more diverse and vary dynamically, posing a significant challenge to sponge acceleration off GPUs.
  
  In this paper, we propose \textsc{Matryoshka}, a novel elastically-parallel technique for the efficient execution of quantum chemistry system with dynamic diversity on GPU. \textsc{Matryoshka} capitalizes on Elastic Parallelism Transformation, a property prevalent in scientific systems yet underexplored for dynamic diversity, to elastically realign parallel patterns with GPU architecture. Structured around three transformation primitives (Permutation, Deconstruction, and Combination), \textsc{Matryoshka} encompasses three core components. The Block Constructor serves as the central orchestrator, which reformulates data structures accommodating dynamic inputs and constructs fine-grained GPU-efficient compute blocks. Within each compute block, the Graph Compiler operates offline, generating high-performance code with clear computational path through an automated compilation process. The Workload Allocator dynamically schedules workloads with varying operational intensities to threads online. It achieves highly efficient parallelism for compute-intensive operations and facilitates fusion with neighboring memory-intensive operations automatically. Extensive evaluation shows that \textsc{Matryoshka} effectively addresses dynamic diversity, yielding acceleration improvements of up to $13.86\times$ (average $9.41\times$) over prevailing state-of-the-art approaches on 13 quantum chemistry systems.
\end{abstract}




\settopmatter{printccs=false, printacmref=false}
\maketitle

\section{Introduction}

  Deep learning, powered by AI infrastructures like GPUs, has sparked transformative revolutions in various AI systems~\cite{ai-alphafold,ai-bert,ai-conv,ai-llama,ai-resnet,ai-transformer}. Concurrently, scientific computing assumes an equally vital role, propelling breakthrough research in various scientific domains~\cite{sc-gb21,sc-gb22,sc-gb23}. However, an intriguing observation lies in the prevailing state where the numerical scientific computing systems predominantly revolve around case-by-case optimizations on CPUs~\cite{cpu-covid,cpu-earthquake,cpu-fft,cpu-stencil-1,cpu-stencil-2}. A noticeable dichotomy emerges as deep learning and scientific computing appear to progress as parallel entities. This intriguing scenario prompts a compelling question: Why the absence of systematic efforts to integrate AI infrastructures like GPUs into scientific computing? 

  This paper aims to unravel this puzzle by delving into quantum chemistry  systems within scientific computing. Quantum Chemistry (QC) ~\cite{QC-review-1,QC-review-2,QC-review-3} is a pivotal scientific discipline that investigates the quantum mechanical properties of atomic and molecular systems. The profound insights gleaned from computational quantum chemistry find extensive applications across diverse industries, including materials science~\cite{QC-material-1,QC-material-2}, pharmaceuticals~\cite{QC-pharmaceuticals-1}, and energy production~\cite{QC-energy-production}.
  
  Notwithstanding its transformative potential, the formidable computational complexity of QC poses a substantial demand on computing resources, thereby impeding progress within the scientific domain. With the booming rise of AI techniques, accelerators, primarily GPUs, have been increasingly employed to expedite QC systems.
  
\begin{figure}
  \includegraphics[width=0.5\textwidth]{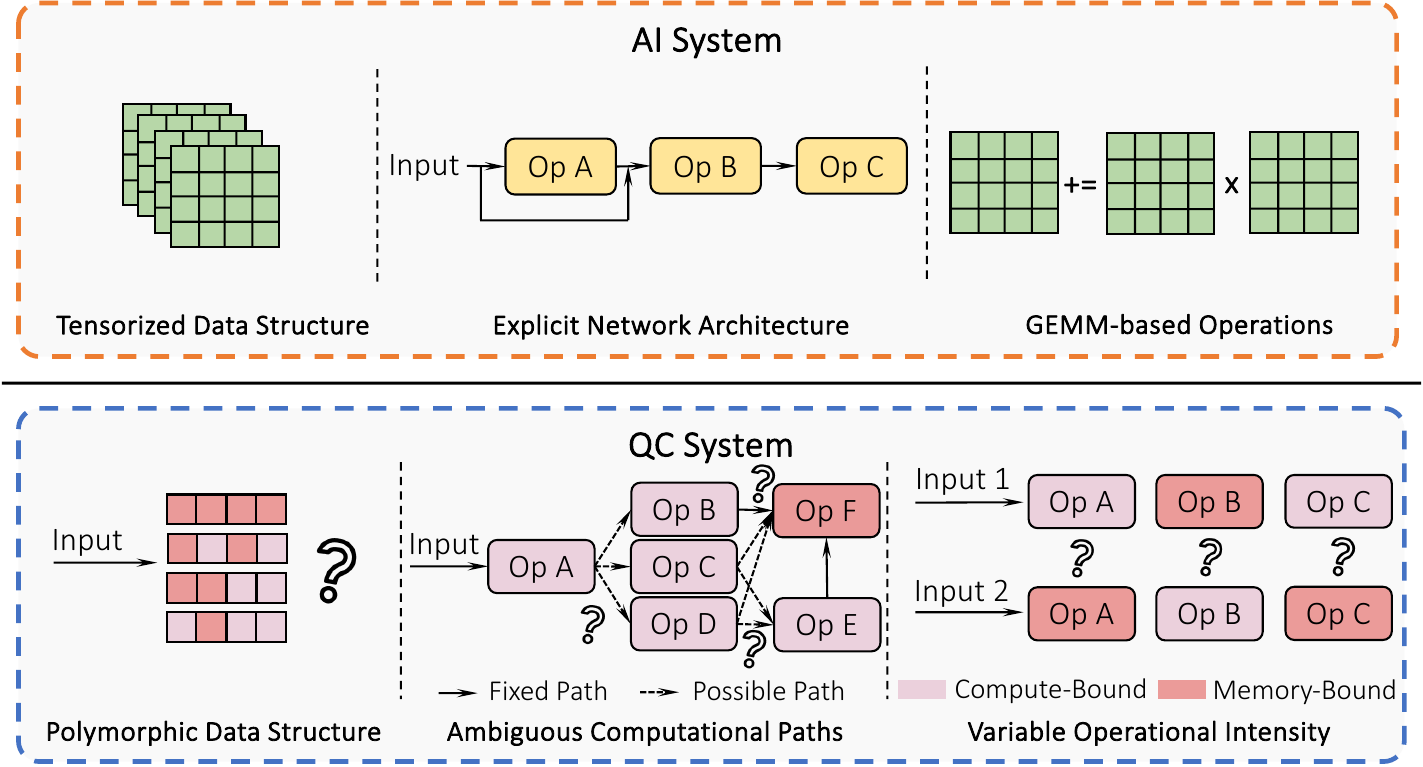}
  \caption{Comparison between AI system and QC system.}
  \label{fig:ai-vs-qc}
  \end{figure}

  While GPU could deliver a promising performance, particularly in General Matrix Multiply (GEMM) operations on tensors predominant in AI systems, it is imperative to recognize that computational patterns in scientific systems are significantly more \textit{diverse}, i.e., a single system is built by various distinct operations with polymorphic data structures. Adding to this complexity, these diverse computational patterns are often varying at compile time or influenced by inputs at runtime, i.e., dynamic diversity. This introduces indiscernible and unpredictable performance bottlenecks in scientific systems, making optimization challenging without sufficient priors, as shown in Figure ~\ref{fig:ai-vs-qc}.
  
  As a fundamental system in scientific computing, QC grapples with the significant challenge of dynamic diversity, evident across three primary misalignments against GPU architecture. These include polymorphic data structures causing divergent execution instructions, ambiguous computational paths contributing to a complex computing landscape at compile time, and variable operational intensity spanning compute-intensive to memory-intensive workloads at runtime.
  
  In the presence of dynamic diversity, efficiently transitioning from CPU-centric to GPU-centric architectures for QC systems remains a substantial challenge, with the majority of current work still rooted in CPU-centric designs. Existing GPU-based QC systems typically perform a fixed optimization strategy on diverse operators known at compile time~\cite{GPU-Ufimtsev-1,GPU-Ufimtsev-2,GPU-Ufimtsev-3,GPU-Ufimtsev-4,GPU-QUICK-1,GPU-QUICK-2,GPU-QUICK-3,GPU-QUICK-4,GPU-Barca}. However, this static parallelism lacks adaptability to dynamic changes in tasks during runtime, leading to a noticeable decrease in GPU utilization. While fine-grained, operator-wise runtime solutions exist, their applicability is constrained to supporting a single module within the QC system. The concatenation of these operators to construct a comprehensive system introduces significant overhead for data layout conversion across diverse operators ~\cite{GPU-Ufimtsev-1,GPU-Ufimtsev-2,GPU-Ufimtsev-3,GPU-Ufimtsev-4}. Furthermore, this approach may lead to performance conflicts when alternative techniques are applied to different operators. Beyond the performance impact, the re-design of operators for mapping to GPU SIMT architecture and the data layout conversion between operators can also incur accuracy loss, which hardly meets the stringent requirements of scientists from a physics perspective ~\cite{GPU-QUICK-1,GPU-QUICK-2,GPU-QUICK-3,GPU-QUICK-4,GPU-Barca}.

  In this paper, we present \textsc{Matryoshka}, a novel elastically-parallel technique for the accurate and efficient execution of quantum chemistry system with dynamic diversity on GPU.
  
  The cornerstone of \textsc{Matryoshka} is Elastic Parallelism Transformation, a property prevalent in scientific systems yet underexplored for dynamic diversity. Most scientific operators have one or more dimensions (we call them EPT-axes), whose computation can be arbitrarily reordered without affecting the result. For instance, in a QC system, contracted basis functions operator $\psi_{a}$ are fixed linear combinations of fundamental basis function $\phi_{ak}$, given by $\psi_{a}(\textbf{r}) = \sum_{k}^{K}{D_{ak}\phi_{ak}(\textbf{r})}$. Leveraging EPT, \textsc{Matryoshka} decouples $\psi_{a}$ along an EPT-axis ($k$ dimension) into multiple compute tiles.
  These tiles serve as parallelism-friendly units of the operator $\psi_{a}$ at various stages, allowing them to be \textit{permuted} in any order, \textit{deconstructed} along the intrinsic EPT-axis, and {\textit{combined} to larger size for realigning an efficient parallelism on GPU elastically}. 
  
  Through the manipulation of compute tiles along EPT-axes, \textsc{Matryoshka} achieves the best of both worlds. It can achieve fine-grained, dense compute tile coverage of diverse operators without incurring layout transformation overhead, while promote highly-adapted parallelism with exhaustive GPU utilization.

  A significant challenge to leverage Elastic Parallelism Transformation is how to capture and address the dynamic characteristics effectively. \textsc{Matryoshka} tackles this challenge by the design of divergence-free Block Constructor, path-explicit Graph Compiler and thread-saturated Workload Allocator, which employs three EPT primitives (Permutation, Deconstruction, and Combination) to   address dynamic diversity in QC systems.
  
  The Block Constructor serves as the central orchestrator. It reformulates versatile data structures immune to dynamic inputs, and generates fine-grained blocks for efficient mapping to GPU SIMT architecture. The Graph Compiler, operating offline, transforms original scientific operators into computation graphs, applies domain-specific expert knowledge distilled from fine-tuned optimizations, and produces highly path-clear yet execution-efficient code through an automated compilation process. Meanwhile, the Workload Allocator combines compute tiles with varying operational intensity online. It rapidly evaluates workload schedule proposals, automatically allocating workloads to threads based on the strategic selection. This results in highly efficient parallelism for compute-intensive operations and facilitates fusion with neighboring memory-intensive operations.
 
  We conduct a comprehensive evaluation of \textsc{Matryoshka} on 13 representative QC systems (Chignolin, DNA, Crambin, etc.), demonstrating a remarkable speedup of up to 5.9x compared to state-of-the-art solutions. Notably, existing state-of-the-art approaches, while preserving original accuracy, have not surpassed the simulation scale of more than 1,000 atoms~\cite{CPU-GAMESS,CPU-PySCF-1,CPU-PySCF-2,CPU-PySCF-3,GPU-QUICK-1,GPU-QUICK-2,GPU-QUICK-3,GPU-QUICK-4}. \textsc{Matryoshka} achieves a breakthrough, maintaining original \textit{ab initio} accuracy while simulating 11,259 atoms using a single GPU in one day (19.5 hrs). 
  
  Having traversed the content thus far, the rationale behind the initial posed question has been revealed: dynamic diversity. In response to this challenge, we introduce \textsc{Matryoshka}. \textsc{Matryoshka} signifies a paradigm shift, ushering in a new era-System4Science.  This innovation lays the foundation for more efficient and scalable scientific computing on AI infrastructure in the context of quantum chemistry systems and beyond.

\section{Background}

\subsection{Quantum Chemistry System}

  The cornerstone of quantum chemistry lies in the elucidation of atomic systems through the resolution of the Schrödinger equation~\cite{QC-equation}. However, the Schrödinger equation proves tractable solely for a finite set of elementary single-electron systems. To probe more intricate systems, the preponderance of computational quantum chemistry methodologies, including the widely employed Hartree-Fock (HF) and density-functional theory (DFT), resort to the self-consistent field (SCF) approximation. This approximation posits that each electron within the system traverses an average potential generated by all other electrons, forming the basis for constructing a QC system~\cite{DFT}.

  The resolution process of a QC system is completed through iterative steps. Commencing with an initial conjecture for the molecular orbitals (MOs), the process entails constructing the Fock matrix employing the extant MOs, which encapsulates contributions from kinetic energy, two-electron repulsion, and an effective potential. Subsequently, the electron density matrix is derived from the occupied MOs, and the total energy of the system is computed. Finally, an evaluation ensues to ascertain the convergence of electronic density and total energy to a stable solution. Should convergence remain elusive, the MOs are updated based on the prevailing electron density, and the entire process is reiterated until convergence is attained.

\subsection{Electron Repulsion Integrals\label{sec:eri}}
  
  \begin{table}
      \caption{Symbol Table in \S ~\ref{sec:eri}}
      \label{tab:symbols}
      \begin{tabular}{ll}
      \hline
          Symbol & Meaning \\
      \hline
          $\psi$ & Contracted basis function \\
          $\phi$ & Fundamental basis function \\
          $\mathbf{a}$, $\mathbf{b}$, $\mathbf{c}$, $\mathbf{d}$ & Angular momentum of basis function\\
          $\mathbf{a}_x$, $\mathbf{a}_y$, $\mathbf{a}_x$ & Components of the angular momentum $\mathbf{a}$ \\
          $K$, $L$, $M$, $N$ & Degree of contraction \\
          $D_{ak}$ & Contraction coefficient \\
          $(\mathbf{a}\mathbf{b}|\mathbf{c}\mathbf{d})$ & Contracted ERI \\
          $[\mathbf{a}\mathbf{b}|\mathbf{c}\mathbf{d}]$ & Fundamental ERI \\
      \hline
      \end{tabular}
  \end{table}

  The computation of two-electron repulsion integrals (ERIs) within a QC system constitutes a significant time consumption, accounting for approximately 95\% of the overall computation time ~\cite{pople1978computation}.

  In quantum mechanics,  a contracted basis function $\psi$ characterizes an electron's spatial position, representing a linear combination of more fundamental basis functions $\phi$. Each $\phi$ shares identical angular momentum, defined by the vector $\textbf{a}$. This vector consists of three integers ($a_{x}$, $a_{y}$, $a_{z}$), with their cumulative sum determining the angular momentum of the fundamental basis function. Mathematically, this relationship is expressed as:
  \begin{equation}
    \psi_{a} = \sum_{k}^{K}{D_{ak}\phi_{ak}}.
  \end{equation}
  Here, the length of the linear combination, denoted as $K$, is termed the degree of contraction, and $D_{ak}$ represents the contraction coefficients.

  Angular momentum serves as a unique identifier for the contracted basis function. Assuming the angular momentum of the four involved contracted basis functions as $\textbf{a}$, $\textbf{b}$, $\textbf{c}$, $\textbf{d}$, the integral between two electrons can be denoted as $(\mathbf{a}\mathbf{b}|\mathbf{c}\mathbf{d})$, which involves four contracted basis functions.

  Given that the contracted basis function is a linear combination of fundamental basis functions, the integral $(\mathbf{a}\mathbf{b}|\mathbf{c}\mathbf{d})$ can be expanded into the following form of basis function quadruple~\cite{HGP}:
  \begin{equation}
  \label{eq:contraction}
      (\mathbf{a}\mathbf{b}|\mathbf{c}\mathbf{d}) = \sum_{k}^{K}{\sum_{l}^{L}{\sum_{m}^{M}{\sum_{n}^{N}{D_{ak}D_{bl}D_{cm}D_{dn}}[\textbf{a}_{k}\textbf{b}_{l}|\textbf{c}_{m}\textbf{d}_{n}]}}}
  \end{equation} 

  To solve $[\mathbf{a}\mathbf{b}|\mathbf{c}\mathbf{d}]$, a recurrence relations algorithm is employed in ERI computations. The key idea is the calculation of $[\mathbf{a}\mathbf{b}|\mathbf{c}\mathbf{d}]$ with larger angular momentum  can be derived from previously computed ones with smaller angular momentum. The algorithm involves two principles: Horizontal Recurrence Relation (HRR), which shifts angular momentum by 
  transforming $[\mathbf{a}\mathbf{b}|\mathbf{c}\mathbf{d}]$ into $[\mathbf{(a+b)}\mathbf{0}|\mathbf{(c+d)}\mathbf{0}]$; and Vertical Recurrence Relation (VRR), which reduces angular momentum at the $\mathbf{a}$ or $\mathbf{c}$. Once all angular momenta at these 4 positions have been reduced to $\mathbf{0}$, the integral can be computed analytically on $[\mathbf{0}\mathbf{0}|\mathbf{0}\mathbf{0}]$. 
  
\subsection{State-of-the-art}

  Recent efforts to accelerate research on large-scale QC systems with ab initio accuracy have pursued two primary approaches~\cite{CPU-GAMESS,GPU-Ufimtsev-1,GPU-Ufimtsev-2,GPU-Ufimtsev-3,GPU-Ufimtsev-4,GPU-Asadchev,GPU-QUICK-1,GPU-QUICK-2,GPU-QUICK-3,GPU-QUICK-4,GPU-Barca}. One approach involves utilizing a greater number of CPU cores for distributed computing acceleration, a solution that consumes vast computational resources and is prohibitively expensive~\cite{CPU-GAMESS}.
  The other focuses on efficiently porting QC systems to GPU architecture using High Performance Computing (HPC) techniques. However, as discussed earlier, the current approaches face significant challenges due to dynamic diversity, which considerably limits the potential of GPUs\cite{GPU-Ufimtsev-1,GPU-Ufimtsev-2,GPU-Ufimtsev-3,GPU-Ufimtsev-4,GPU-Asadchev,GPU-QUICK-1,GPU-QUICK-2,GPU-QUICK-3,GPU-QUICK-4,GPU-Barca}. 
  To the best of our knowledge, the industry-recognized state-of-the-art approaches  maintaining original ab initio accuracy have yet to surpass the scale of simulating more than 1,000 atoms~\cite{GPU-QUICK-1,GPU-QUICK-2,GPU-QUICK-3,GPU-QUICK-4,GPU-Barca}. Consequently, the design at the system level to efficiently support larger-scale QC systems with fewer computational resources emerges as a crucial and challenging endeavor.

\section{Key Challenges and Insights}

\subsection{Dynamic Diversity}

  The existence of dynamic diversity presents a notable challenge, creating a significant misalignment between scientific systems and AI infrastructure with SIMT architecture. Over an extended period, the development and optimization of scientific systems have predominantly revolved around CPU architecture with serial computation logic. Consequently, these systems have not fully harnessed the advantages offered by cutting-edge AI infrastructure, such as GPUs renowned for their high performance. In QC systems, this limitation becomes particularly pronounced across three key dimensions.

  \textbf{Polymorphic data structures.} Unlike the prevalent tensors with uniform data structures found in AI systems, a QC system is marked by a diverse array of data structures, including basis function, basis function pair, and basis function quadruple, to describe integrals of different input atoms. This diversity introduces a complex data organization and distinct execution instructions, leading to two key challenges. 
  
  Firstly, the involvement of multiple data structures escalates the memory cost, reaching up to the fourth power of basis functions. However, given the relatively constrained GPU memory compared to CPUs, accommodating large-scale QC systems becomes challenging. Secondly, the existence of distinct execution instructions for these diverse data structures often leads to warp divergence. This phenomenon emerges when threads within a warp deviate from a uniform execution path, resulting in only one thread being active in a warp. Consequently, this divergence triggers the serialization of execution and low throughput on the GPU.

  \textbf{Ambiguous computational paths.} In AI systems, model structures are designed hierarchically, establishing explicit and well-defined computation paths through interconnected layers. In contrast, the computation paths in QC systems are highly diverse and non-unique, resulting in an intricate and continually-changing computational landscape for the entire system. This diversity introduces two significant challenges. 
  
  Firstly, the complex computation paths bring forth numerous high-precision floating-point operations, rapidly depleting limited register resources. This, in turn, leads to register spilling and subsequent performance degradation. Secondly, in existing work, all candidate computational paths are often manually coded and individually optimized, resulting in suboptimal efficiency and poor scalability when applied to larger systems. Adding to the complexity, for a specific computational path, its cost is often dynamic. This dynamism arises because the computational workload is influenced by the input at each recursive entrance during runtime. This dynamic characteristic transforms each computation path into a highly serial and unpredictable process.
  
  \textbf{Variable operational intensity.} GEMM operation serves as a fundamental and pivotal component within the domain of AI systems. Renowned for its high parallelizability, GEMM is particularly well-suited for hardware acceleration, notably on GPUs.

  In stark contrast, the operations involved in QC systems deviate markedly from the prevalent homogeneity observed in AI systems. These operations display substantial variability across different types of atom, ranging from compute-intensive to memory-intensive. In particular, the locations where these operations are bound can dynamically shift. This implies that a specific operation may transition from being compute-intensive to memory-intensive based on varying inputs, resulting in notable idle periods that only become discernible at runtime.

\subsection{Elastic Parallelism Transformation}

  \begin{figure}
      \includegraphics[width=0.48\textwidth]{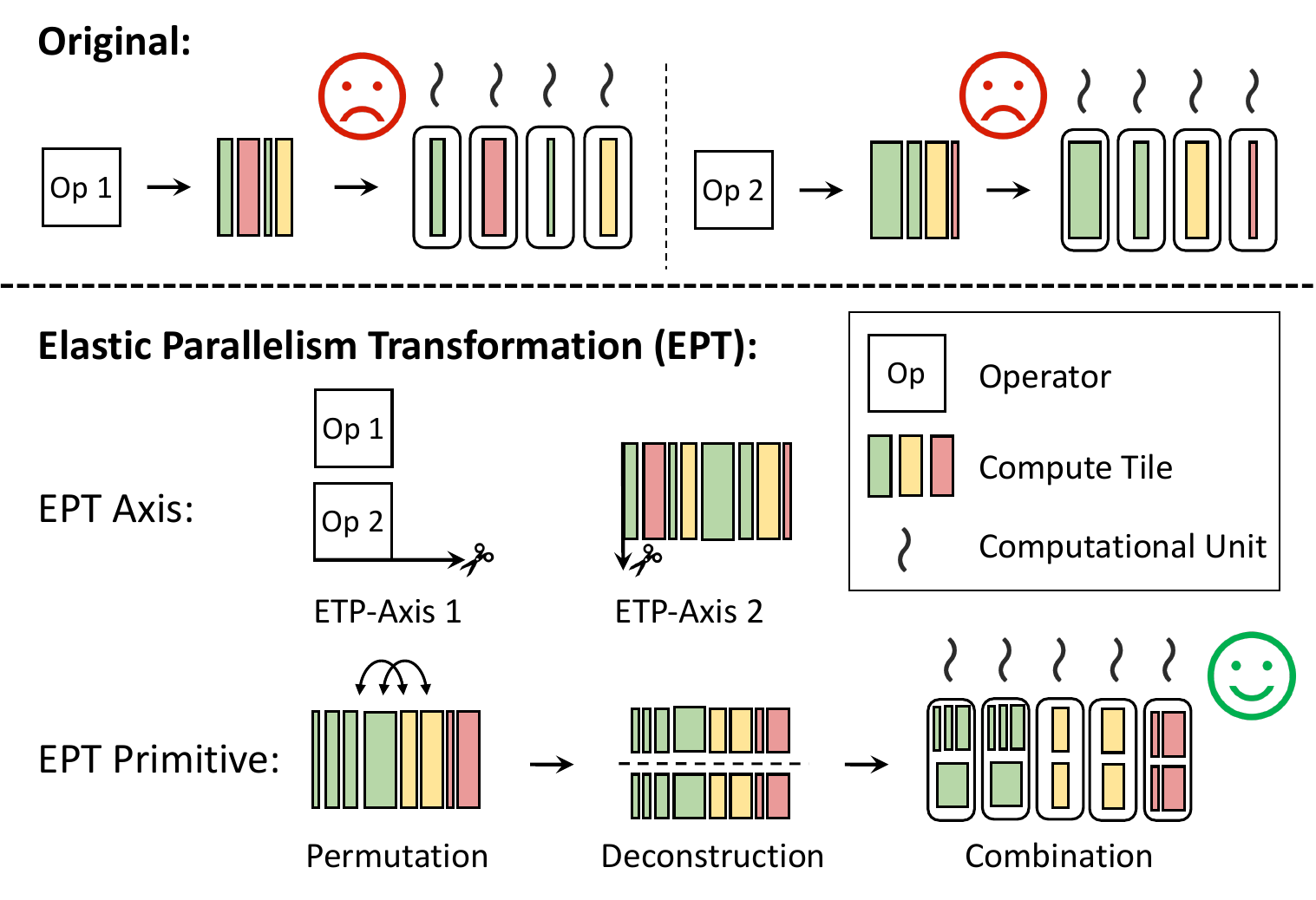}
      \caption{Illustration of Elastic Parallelism Transformation. Through the identification of EPT-axes and the application of EPT primitives, we can achieve fine-grained, dense compute tile coverage of diverse operators without incurring layout transformation overhead, while promote GPU utilization.}
      \label{fig:ept}
  \end{figure} 
  
  As shown in Figure ~\ref{fig:ept}, Elastic Parallelism Transformation stands as the cornerstone insight employed by \textsc{Matryoshka} to address dynamic diversity within QC systems, contributing to heightened GPU-efficient parallelism without incurring any data layout conversion costs.

  In QC systems, each operator typically involves multiple reduction dimensions. \textsc{Matryoshka} designates these dimensions as EPT-axes if and only if all computations on this axis are commutative and associative. The commutative and associative property ensures mathematically equivalent computation transformations on the original operators, allowing for the random shuffling of compute tiles along the EPT-axis to achieve a more elastic parallel pattern in any order.
  
  Along the EPT-axis, \textsc{Matryoshka} introduces the design of compute tile—a fragment of an operator split in an adjustable size, which serves to compose a parallelism-friendly unit for efficient computation. For instance, in a QC system where contracted basis functions ($\psi_{a}$) are linear combinations of fundamental Gaussians ($\phi_{ak}$), EPT recognizes that $\psi_{a}$ has one EPT-axis ($k$ dimensions), allowing the operator to be split into at most $k$ independent $\phi_{ak}$ compute tiles along this axis.
  
  Based on the introduced concepts of EPT-axis and compute tiles, EPT defines three pivotal transformation primitives. Each primitive works on a specific EPT-axis, orchestrating precise transformations within operators by manipulating nested compute tiles:
  \begin{enumerate}[label=(\arabic*), nosep, leftmargin=*]
      \item \textbf{Permutation}: Along an EPT-axis, a portion or the entirety of compute tiles can rearrange their calculation order.
      \item \textbf{Deconstruction}: If a compute tile retains an EPT-axis, it can undergo further split into more sub-compute tiles.
      \item \textbf{Combination}: Sub-compute tiles with the same EPT-axis can also be nested into a larger compute tile.
  \end{enumerate} 
  
  Through the strategic application of these three primitives, dynamic diversity within QC systems can be effectively addressed by elastically aligning the nested compute tiles with the GPU architecture.

\section{Design Overview}

\begin{figure}
      \includegraphics[width=0.35\textwidth]{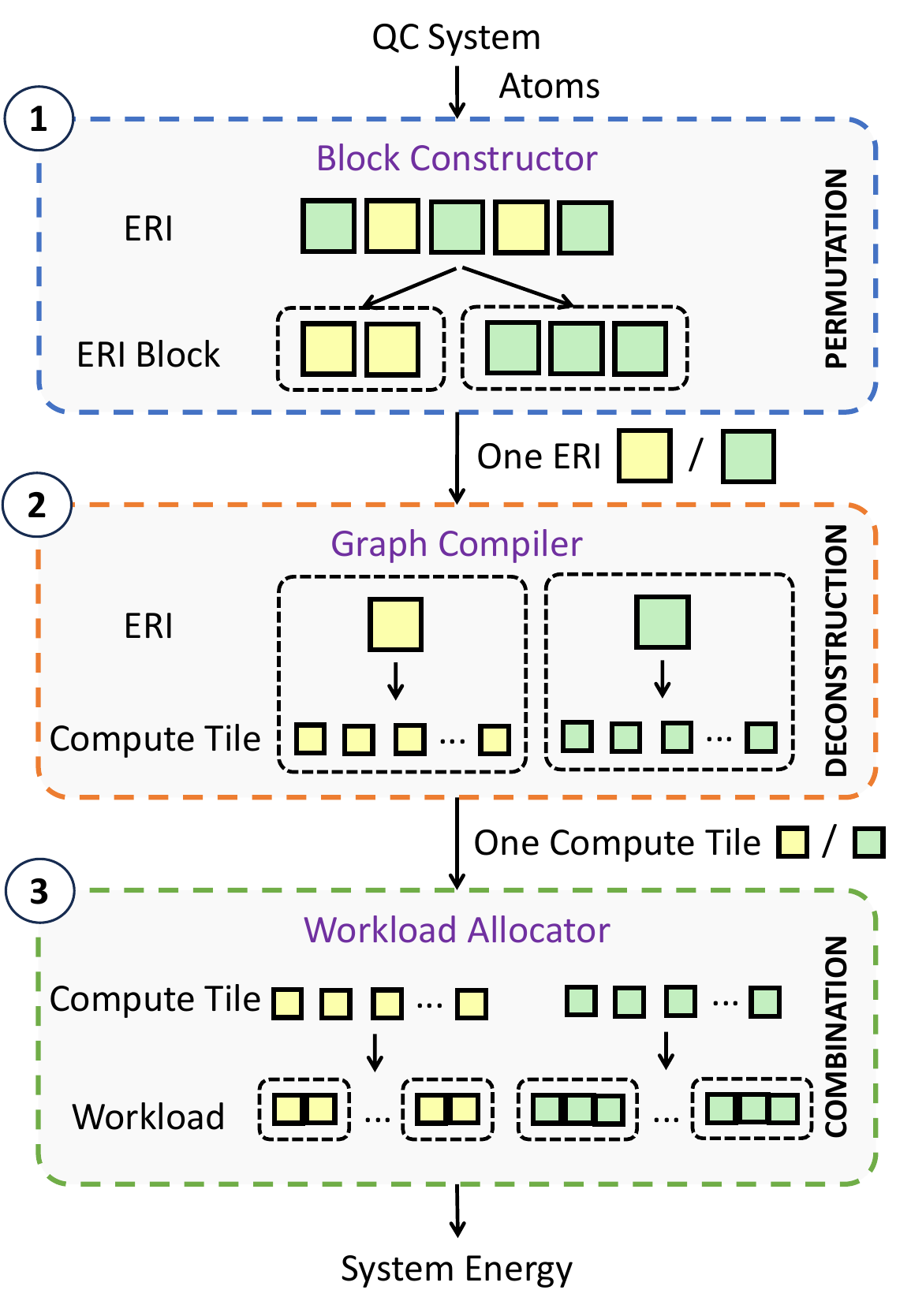}
      \caption{Overview of \textsc{Matryoshka}.}
      \label{fig:overview}
  \end{figure}

  Figure~\ref{fig:overview} shows an overview of \textsc{Matryoshka}, which consists of three core components: divergence-free Block Constructor, path-explicit Graph Compiler, and thread-saturated Workload Allocator.

  The Block Constructor serves as the central orchestrator  for the clustering of ERIs and the construction of ERI blocks with consistent execution instructions, which leverages the Permutation EPT primitive. It reformulates versatile data structures resilient to dynamic inputs and generates fine-grained ERI blocks tailored for efficient mapping onto GPU SIMT architecture (\S ~\ref{sec:block}).

  In an offline manner, the Graph Compiler produces highly path-clear yet execution-efficient code through an automated compilation process by utilizing the Deconstruction EPT primitive. The domain-specific expert knowledge is also distilled into the Compiler to guide an efficient path-search algorithm (\S ~\ref{sec:compiler}). 
  
  Employing the Combination EPT primitive, the Workload Allocator dynamically schedules ERI workloads in response to their varying operational intensities online. Rapid evaluation of workload schedule proposals and automatic assignment of workloads to threads are key features, resulting in highly efficient parallelism for compute-intensive operations and facilitating fusion with adjacent memory-intensive operations (\S ~\ref{sec:allocator}).
  
\section{Block Constructor\label{sec:block}}

  In contrast to the uniform data structures, such as tensors, that facilitate batch processing in AI systems, QC systems present a series of distinctive challenges of Dynamic Diversity due to their reliance on polymorphic data structures. 

  The primary challenge is the accelerated depletion of computational and memory resources. As detailed in \S ~\ref{sec:eri}, the ERI computation centers around a specific data structure—the basis function quadruple $(\mathbf{a}\mathbf{b}|\mathbf{c}\mathbf{d})$. In a QC system with $N$ basis functions, the computation involves $O(N^4)$ ERIs of diverse types. As the size of QC systems scales up, the total number of ERIs reaches an overwhelming order of magnitude. Managing such a voluminous set is not only time-consuming but also places an immense demand on memory resources.

  The second challenge is the issue of warp divergence during GPU computations. In QC systems, basis functions sharing the same angular momentum belong to the same class with consistent execution instructions. However, when a substantial number of basis functions are sent to the GPU for computation, threads within one warp may compute ERIs belonging to different classes. This divergence necessitates different threads executing different instructions, resulting in severe warp divergence.

  Moreover, the distribution of these quadruples is inherently stochastic and undergoes dynamic changes with different QC systems. This dynamic nature implies that preprocessing these ERIs at compile time is unfeasible; instead, it must be conducted at runtime.

  In \textsc{Matryoshka} we propose the Block Constructor which constructs basis function quadruple blocks in a streaming manner to handle the challenges of Dynamic Diversity caused by polymorphic data structures.
  
  The Block Constructor is conceived based on two key insights, employing the initial EPT primitive—Permutation. Firstly, \textsc{Matryoshka} discerns that each basis function quadruple inherently possesses an EPT-axis of basis function pairs. Along this EPT-axis, $(\mathbf{a}\mathbf{b}|\mathbf{c}\mathbf{d})$ can be derived from the permutation of two pairs of basis functions, namely $(\mathbf{a}\mathbf{b}|$ and $|\mathbf{c}\mathbf{d})$. This realization indicates that there is no necessity to pre-construct all basis function quadruples; rather, constructing all basis function pairs and permuting them as required during computational proceedings suffices. This approach significantly reduces the memory cost of $(\mathbf{a}\mathbf{b}|\mathbf{c}\mathbf{d})$ from $O(N^4)$ to $O(N^2)$. 
  Secondly, acknowledging the independence of each ERI computation within a basis function quadruple, \textsc{Matryoshka} identifies an additional EPT-axis of ERIs. Leveraging the Permutation primitive, the ERI computation can be permuted to group ERIs belonging to the same class together. This flexible adjustment allows for the batch processing of ERIs with consistent execution instructions within a single warp.

  As shown in Figure ~\ref{fig:block-constructor}, the Block Constructor takes two stages to complete the construction:
  \begin{figure}
      \centering
      \includegraphics[width=0.48\textwidth]{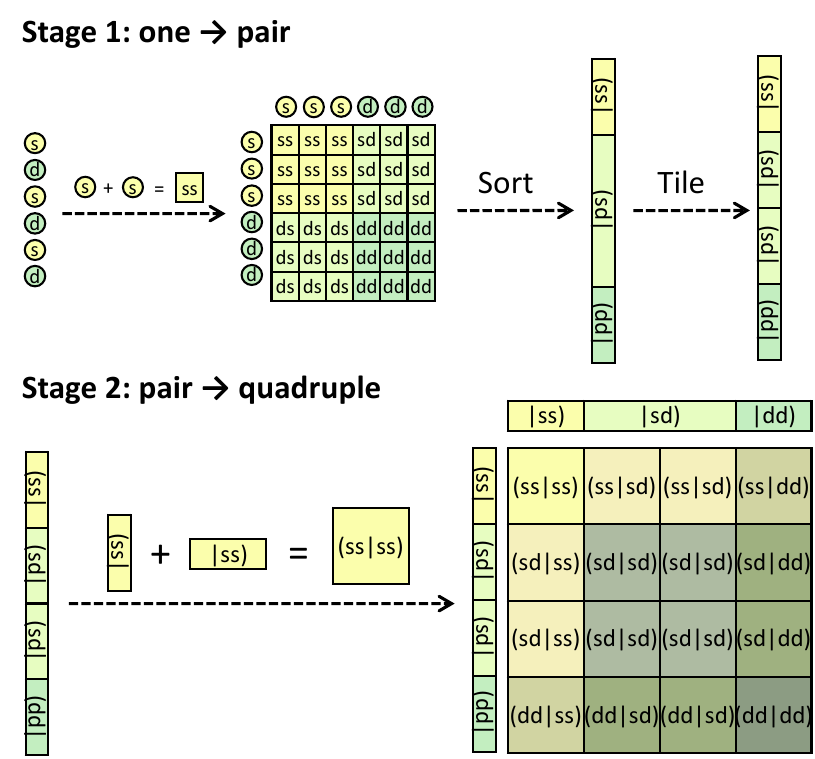}
      \caption{Block Constructor's two-stage construction process. Circles, rectangles, and squares represent basis function individuals, pairs, and quadruples respectively.}
      \label{fig:block-constructor}
  \end{figure}

  \textbf{Stage 1: basis function $\rightarrow$ basis function pair.} In this stage, the Block Constructor constructs an exhaustive pairing of all basis functions. Subsequently, these pairs are sorted in ascending order based on their angular momentum. To optimize data locality, we further segment these basis function pairs into multiple tiles. Notably, to guarantee that the resulting ERIs from the combination of these tiles fall within the same ERI class, the tiling process is exclusively performed within groups of basis function pairs belonging to the same class.

  \textbf{Stage 2: basis function pair $\rightarrow$ basis function quadruple.} In this stage, the Block Constructor permutes the basis function pair tiles to form blocks of basis function quadruples. Should a tile contains $M$ pairs, the resulting block will be the size of $M^2$, enabling the parallel computation of $M^2$ ERIs. These blocks of ERIs become the fundamental units of computation which share no data dependencies with each other. As a result, different ERI blocks can be assigned to different GPU streams for concurrent execution.

  Through the design of the Block Constructor, \textsc{Matryoshka} successfully reduces the memory cost of $(\mathbf{a}\mathbf{b}|\mathbf{c}\mathbf{d})$ from $O(N^4)$ to $O(N^2)$, thereby freeing up a significant amount of memory space. Simultaneously, this design effectively alleviates the warp divergence problem, significantly enhancing the parallelism of ERI computations.

\section{Graph Compiler\label{sec:compiler}}

  After enhancing the efficiency of batching various ERIs with the Block Constructor, this subsection will concentrate on the computation inside a single ERI. As introduced in \S~\ref{sec:eri}, recurrence relations algorithm is used to compute a single ERI $(\mathbf{a}\mathbf{b}|\mathbf{c}\mathbf{d})$ in QC systems. It first reduces the angular momentum on all four positions to $\mathbf{0}$ using these relations and then computes the ERI $(\mathbf{0}\mathbf{0}|\mathbf{0}\mathbf{0})$ analytically. 

  The dynamic diversity poses two main challenges in this procedure. First, the ERI computation involves diverse complex arithmetic operations with high-precision floating-point numbers, including division, modulo, exponentiation, etc. These operations necessitate a substantial allocation of registers per thread, often leading to severe register spilling. Second, a single ERI can give rise to diverse unclear computational paths during the recurrence process. This is because the order of applying the recurrence relations can change, yet all paths can yield the same result. Moreover, different computational paths have dynamic computational costs, which makes pinpointing the most efficient computational path a complex task. These computational costs are not determined until the recurrence process is actually executed.
   
  In \textsc{Matryoshka} we introduce the Graph Compiler, which automatically generates an optimized kernel for each ERI. The Graph Compiler provides four stages to address the above challenges.
   
  \textbf{Stage 1: Computation Deconstruction.}  First, the Graph Compiler identifies the EPT axis existed in the dimension of $(\mathbf{a}\mathbf{b}|\mathbf{c}\mathbf{d})$ according to the principle of EPT. In Equation ~\ref{eq:contraction}, $(\mathbf{a}\mathbf{b}|\mathbf{c}\mathbf{d})$ is calculated from the sum of all $[\mathbf{a}_k\mathbf{b}_l|\mathbf{c}_m\mathbf{d}_n]$. Since different $[\mathbf{a}_k\mathbf{b}_l|\mathbf{c}_m\mathbf{d}_n]$ share no data dependencies, they are naturally commutative and associative. Drawing on the Deconstruction EPT primitive, the Graph Compiler then deconstructs the computation of $(\mathbf{a}\mathbf{b}|\mathbf{c}\mathbf{d})$ into $K*L*M*N$ $[\mathbf{a}\mathbf{b}|\mathbf{c}\mathbf{d}]$ (where $K$, $L$, $M$, and $N$ represent the contraction degrees of the four contracted basis functions involved in the ERI computation). As a result, the basic computational workload for each thread is reduced by a factor of $K*L*M*N$, which markedly mitigates register spilling.

  \begin{figure}
      \centering
      \includegraphics[width=0.48\textwidth]{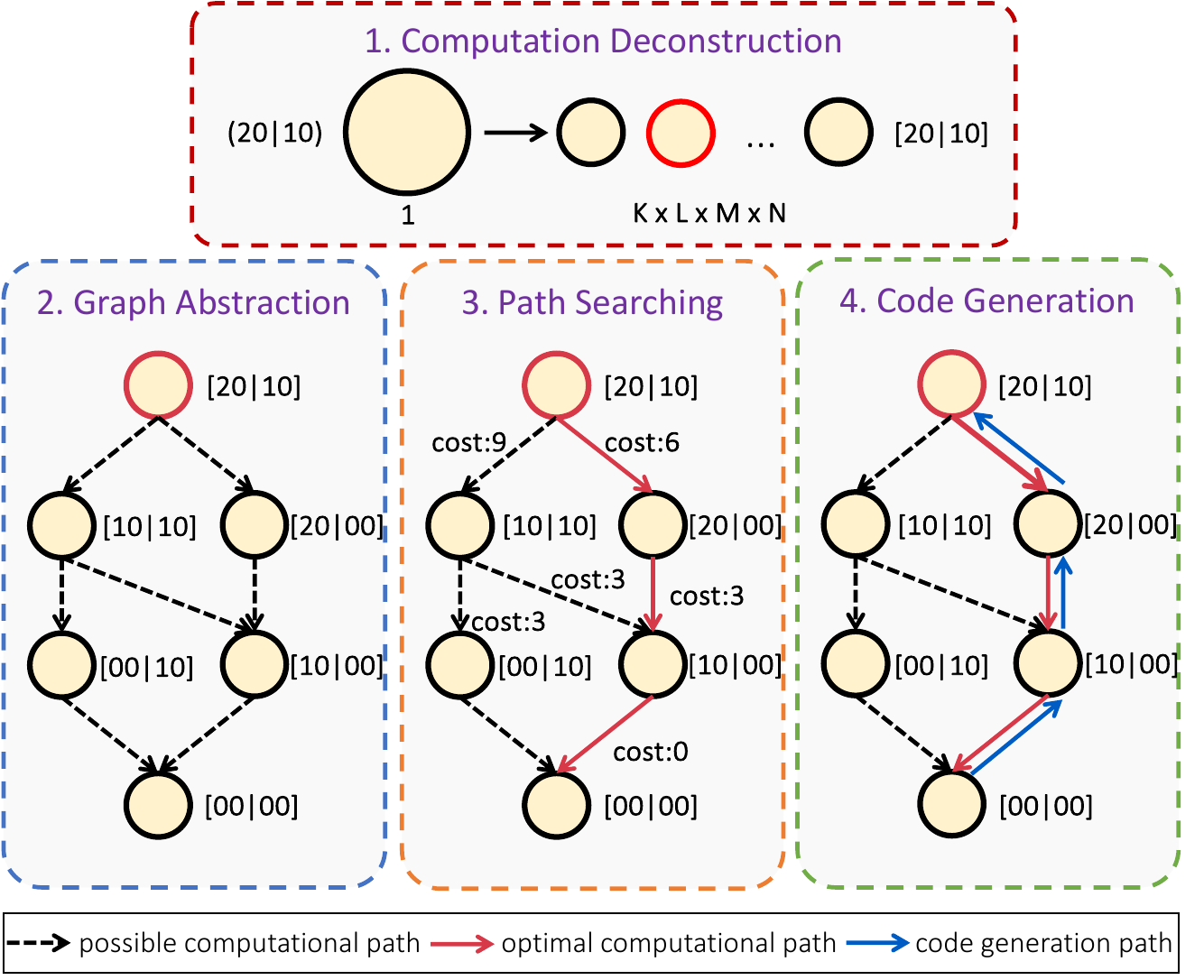}
      \caption{Graph Compiler's four-stage workflow on optimizing $[\mathbf{2}\mathbf{0}|\mathbf{1}\mathbf{0}]$.}
      \label{fig:graph-compiler}
  \end{figure}
    
  \textbf{Stage 2: Graph Abstraction.} The Graph Compiler first abstracts the recurrence process involved in computing each ERI into a \textit{Directed Acyclic Graph} (DAG). Within this DAG, each node represents an intermediate result generated during the recursion process and each edge represents that one intermediate result can be derived from another. The direction of the edge reflects the direction of derivation. This abstraction simplifies the complex recurrence process into a format that is more accessible for computational handling, serving as the foundational basis for the following stages.
    
  \textbf{Stage 3: Path Searching.} Drawing on the recurrence process of ERI computation, we observe that the computational cost of ERI mainly depends on two factors: (1) The length of the computational path; (2) The degree to reuse intermediate results. The first is because the longer the computational path is, the amount of computation naturally grows. The second is because thanks to the recurrence relations, many intermediate results in the previous recurrence steps actually can be reused in the following recurrence steps, which can greatly reduce the amount of computation.
    
  Based on these observations, we conclude three guiding principles for the Graph Compiler to identify an optimized computational path:
  \begin{enumerate}[label=(\arabic*), nosep, leftmargin=*]
      \item Strive for $\mathbf{0}$ angular momentum whenever possible.
      \item Prioritize the reuse of intermediate results from previous computations.
      \item Minimize the generation of new intermediate results.
  \end{enumerate}
  The first principle is designed to terminate the recurrence process as soon as possible. The intent of the other two principles is to diminish computational efforts by preventing unnecessary recalculations and limiting the introduction of new calculations.
  
  Based on these three principles, We design a greedy algorithm in \textsc{Matryoshka} to search for an optimized path on DAG. As Algorithm~\ref{algo:graph-compiler} shown, we express this algorithm as a function to determine the appropriate recurrence position at each recurrence step. From the Line 4, the algorithm iterates through each possible recurrence positions and greedily selects the one with the lowest computational cost. In Line 8, the computational cost is calculated based on the three guiding principles. Particularly, we use a hyperparameter $\lambda$ to balance the first principle with the other two.      
  \begin{algorithm}\small
      \caption{Greedy Path Searching}
      \label{algo:graph-compiler}
      \begin{algorithmic}[1]
      \Function{FindOptimalPosition}{$positions$, $\lambda$}
      \State $minC \leftarrow \infty$
      \State $optPos \leftarrow \text{NULL}$
      \For{each $pos$ in $positions$}
      \State $r \leftarrow \text{count of reused results at } pos$
      \State $n \leftarrow \text{count of new results at } pos$
      \State $a \leftarrow \text{value of angular momentum at } pos$
      \State $cost \leftarrow (n - r) + \lambda * a$
      \If{$cost < minC$}
      \State $minC \leftarrow cost$
      \State $optPos \leftarrow pos$
      \EndIf
      \EndFor
      \State \textbf{return} $optPos$
      \EndFunction
      \end{algorithmic}
  \end{algorithm}
    
  \textbf{Stage 4: Code Generation.} The Graph Compiler accepts the optimized computational path as input, interpreting it as a sequence of nodes arranged by topological order on a DAG. It then reverses this sequence and begins to generate computation codes starting from the base case of recurrence, which is $[\mathbf{0}\mathbf{0}|\mathbf{0}\mathbf{0}]$. After determining the value of ERI $[\mathbf{0}\mathbf{0}|\mathbf{0}\mathbf{0}]$ analytically, the Graph Compiler proceeds to generate code for computing the nodes sequentially, working its way up to the desired target ERI $[\mathbf{a}\mathbf{b}|\mathbf{c}\mathbf{d}]$.

  It is worth noting that the Graph Compiler handles the entire task of kernel generation at compile time, resulting in no overhead during runtime. In addition to generating optimized kernels, the automation of the Graph Compiler equips \textsc{Matryoshka} with outstanding scalability for QC systems of any size. This eliminates the need for manual enumeration and case-by-case optimization across different ERI classes.

\section{Workload Allocator\label{sec:allocator}}

\begin{figure}
      \centering
      \includegraphics[width=0.40\textwidth]{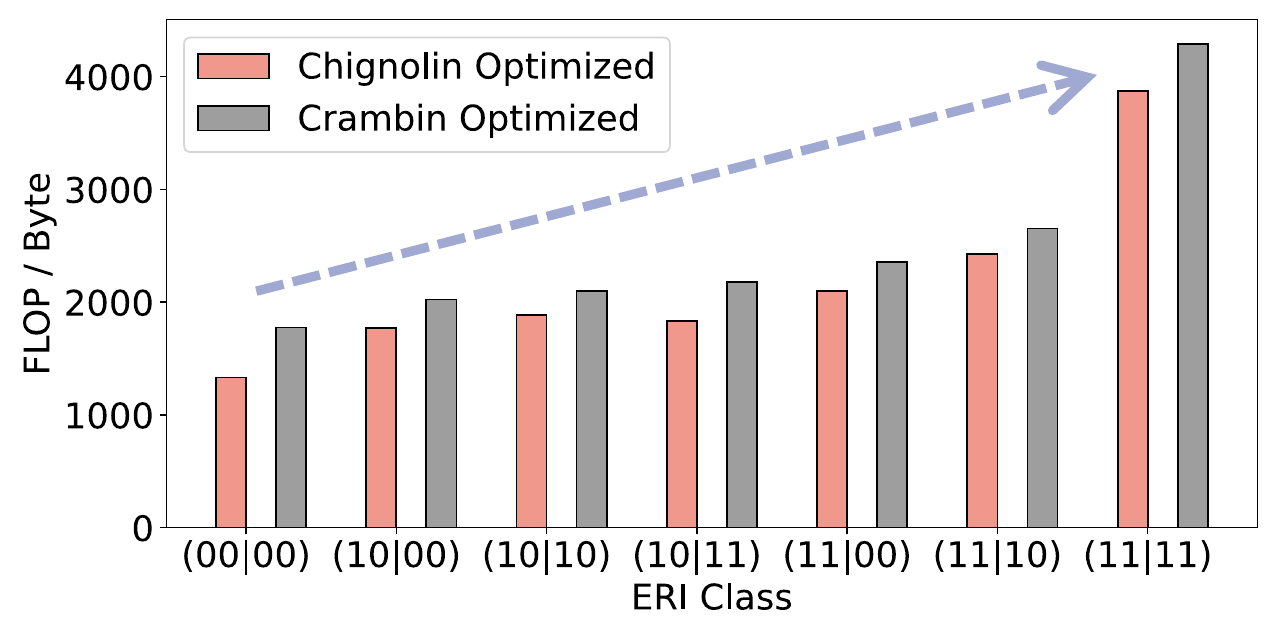}
      \caption{OP/B trends in Chignolin and Crambin. An upward trend is observed between OP/B and angular momentum.}
      \label{fig:workload-allocator-OP2B}
  \end{figure}

  While the Graph Compiler offers optimized kernels for $[\mathbf{a}\mathbf{b}|\mathbf{c}\mathbf{d}]$ computations, it is noteworthy that different kernels for $[\mathbf{a}\mathbf{b}|\mathbf{c}\mathbf{d}]$ exhibit varying operational intensities, spanning from compute-intensive to memory-intensive. In this context, we conduct an analysis of the compute-to-memory ratio (OP/B) for various $[\mathbf{a}\mathbf{b}|\mathbf{c}\mathbf{d}]$ operations within Chignolin, a widely-used QC system composed of an artificial mini-protein. As depicted in Figure ~\ref{fig:workload-allocator-OP2B}, an increase in angular momentum corresponds to a rise in the OP/B of $[\mathbf{a}\mathbf{b}|\mathbf{c}\mathbf{d}]$. This trend arises from the fact that in each ERI computation, the number of memory operations tends to remain stable, while the number of computational operations grows with increasing angular momentum. The multitude of ERI computations with diverse angular momenta leads to a frequent transition in operational intensity between memory-intensive and compute-intensive scenarios.
  
  More complicated, however, is the dynamic nature of operational intensity for each kernel, where the bound position can vary with inputs. This dynamism implies a limited potential to pre-determine the bound bottleneck at compile time, often resulting in the generation of idle bubbles in the computing pipeline and leading to suboptimal GPU utilization at runtime.
  
\begin{figure}
      \includegraphics[width=0.40\textwidth]{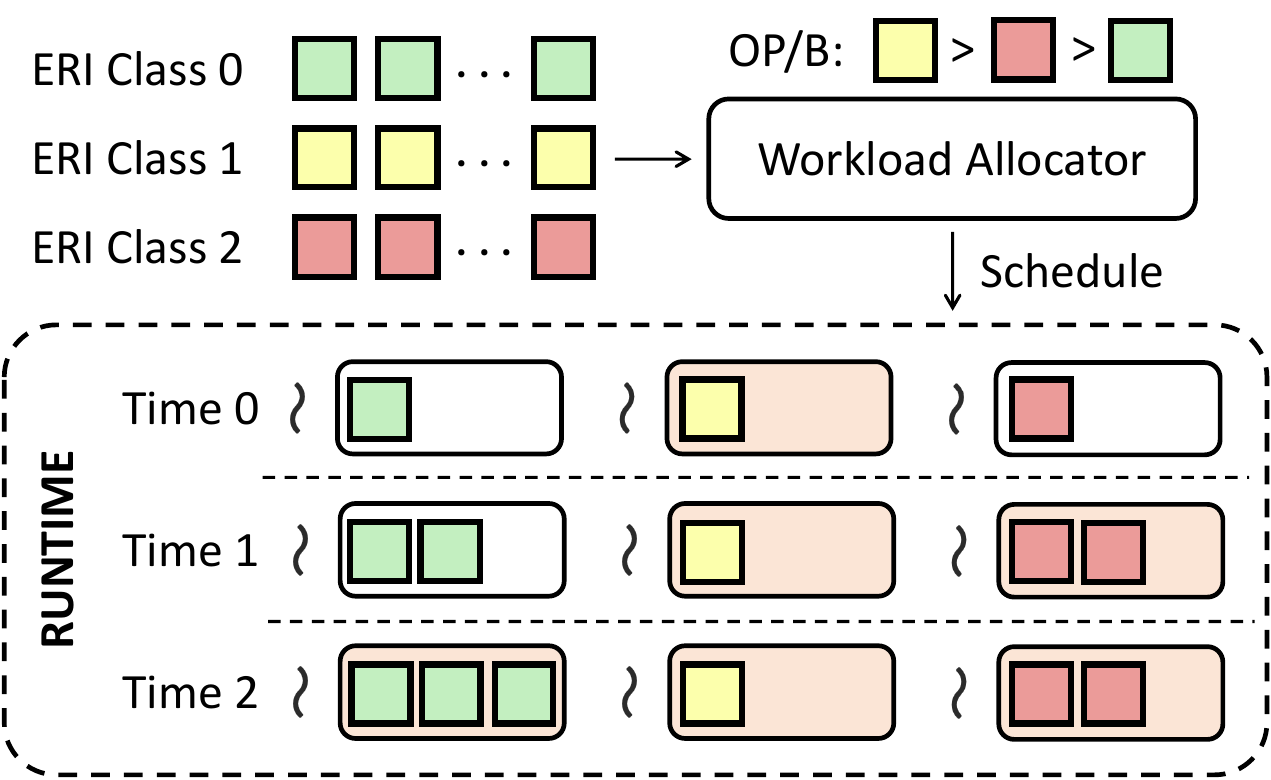}
      \caption{Workload Allocator matches threads to their ideal workload with varying operational intensity. At time 0, each thread is allocated a basic computational unit. Over time, the Workload Allocator progressively assigns additional tasks to the threads until they are saturated (orange-colored).}
      \label{fig:workload-allocator}
  \end{figure}

  As shown in Figure~\ref{fig:workload-allocator}, \textsc{Matryoshka} introduces the Workload Allocator—a key component designed to automatically allocate workloads to each thread efficiently, exploiting both the computational power and memory bandwidth of the GPU. Guided by the Combination EPT primitive, the Workload Allocator initially  expands the $[\mathbf{a}\mathbf{b}|\mathbf{c}\mathbf{d}]$ into larger dependency-free compute tiles with varying degrees of combination first. Then two allocation principles are presented to achieve an efficient overlapping of computation and memory access by managing these compute tiles. 
  \begin{enumerate}[label=(\arabic*), nosep, leftmargin=*]
      \item \textbf{For memory-intensive operations.} The Workload Allocator tends to allocate a larger workload per thread. Since computation is constrained by memory access, assigning more computational workloads can help mitigate idle bubbles caused by memory access latency.
      \item \textbf{For compute-intensive operations.} Adding more computational workload may not be advantageous, as it could already exceed each thread's computational capacity. In such cases, distributing the workload across more threads can help reduce computational latency. Simultaneously, the additional memory access associated with this approach can be mitigated by the underutilized memory bandwidth.
  \end{enumerate}
  
  Based on the two allocation principles, we design an auto-tuning framework for efficiently scheduling workloads with varying operational intensities within the Workload Allocator. As shown in Algorithm~\ref{algo:workload-allocator}, the workload of each thread is allocated as a $[\mathbf{a}\mathbf{b}|\mathbf{c}\mathbf{d}]$ initially in Line 1-2. From the Line 3, the Workload Allocator systematically augments the workload for each ERI class by merging more basic computational units into a bigger one. The criterion for continuing this expansion is contingent upon observing improvements in the wall time of ERI calculations (Line 7-13). At last, the Workload Allocator converges, and each thread aligns with their suitable workload for diverse $[\mathbf{a}\mathbf{b}|\mathbf{c}\mathbf{d}]$.

  \begin{algorithm}\small
      \caption{Auto-tuning Framework}  
      \label{algo:workload-allocator}
      \begin{algorithmic}[1]
      \For{each $cls$ in $ERI\ Classes$}
      \State $workload[cls] \leftarrow [\mathbf{a}\mathbf{b}|\mathbf{c}\mathbf{d}]$
      \EndFor
      \State $improved \leftarrow \text{true}$  
      \While{$improved$}
      \State $improved \leftarrow \text{false}$
      \For{each $cls$ in $ERI\ Classes$}  
      \State $t1 \leftarrow \Call{Time}{cls}$  
      \State $workload[cls] \leftarrow \Call{Combine}{workload[cls]}$  
      \State $t2 \leftarrow \Call{Time}{cls}$  
      \If{$t2 < t1$}  
      \State $improved \leftarrow \text{true}$  
      \Else 
      \State $workload[cls] \leftarrow \Call{Revert}{workload[cls]}$  
      \EndIf 
      \EndFor
      \EndWhile
      \end{algorithmic}
  \end{algorithm}

  Notably, the Workload Allocator seamlessly integrates with ongoing computations at runtime, introducing minimal additional overhead. Furthermore, the Workload Allocator efficiently executes reduction across diverse threads by enabling each thread to employ atomic operations directly. This strategic approach is based on the observation that update positions from different threads exhibit relative sparsity, presenting an opportunity to minimize write conflicts among threads.

\section{Evaluation}

  In this section, we conduct a comprehensive set of experiments to demonstrate the effectiveness of \textsc{Matryoshka} from various perspectives. Specifically, we begin by evaluating the correctness of \textsc{Matryoshka} with Elastic Parallelism Transformation against state-of-the-art approaches on five classical QC systems in \S ~\ref{sec:exp-correctness}. Subsequently, in \S ~\ref{sec:exp-component} we assess each \textsc{Matryoshka} component individually: \S ~\ref{sec:exp-breakdown} shows the performance breakdown of \textsc{Matryoshka}, \S ~\ref{subsec:block-constructor} examines the Block Constructor's role in handling warp divergence, \S ~\ref{subsec:graph-compiler} validates the Graph Compiler's efficacy in mitigating register spilling and optimizing computational paths, and \S ~\ref{subsec:workload-allocator} emphasizes the impact of the Workload Allocator in finely scheduling workloads to alleviate varying operational intensity. The scalability of \textsc{Matryoshka} is explored in \S ~\ref{sec:scalability}, and finally, in \S ~\ref{sec:end-to-end}, we evaluate the end-to-end simulation performance of \textsc{Matryoshka} on both A100 and A6000 GPUs across six larger representative QC systems.  In summary, our results show that:
  \begin{itemize}[nosep, leftmargin=*]
      \item The mathematical rigor of \textsc{Matryoshka} is firmly established through Elastic Parallelism Transformation. From a physicist's stringent viewpoint, the accuracy of \textsc{Matryoshka} can be deemed practically error-free (within $10^{-3}$~\cite{ufimtsev2008quantum,helgaker2008quantitative}), with errors no more than $10^{-5}$ on five representative QC systems in comparison to state-of-the-art approaches.
      \item With Elastic Parallelism Transformation, \textsc{Matryoshka} addresses dynamic diversity effectively. Specifically, the Block Constructor shows a $4.7\times$ improvement, the Graph Compiler a $2.3\times$ improvement, and the Workload Allocator a $4.5\times$ improvement across six QC systems on average.
      \item \textsc{Matryoshka} outperforms the previous state-of-art works and achieves up to $13.86\times$, $9.56\times$, and $4.82\times$ speedup over Libint, PySCF, and QUICK respectively. Besides, it also achieves a breakthrough, maintaining original ab initio accuracy while simulating 11,259 atoms for 99 iterations using a single GPU in one day (19.5 hrs).
  \end{itemize}
  
\subsection{Experimental Setup}

  \textit{Platforms.} We conduct our evaluation on two platforms. Platform A comprises an AMD EPYC 7V13 processor with 24 physical cores, and an Nvidia A100 80GB PCIe GPU. The A100 GPU could provide 1,935 GB/s memory bandwidth and 9.7 TFLOPs for FP64 operations. Platform B consists of an AMD EPYC 7742 processor with 64 physical cores, and four Nvidia A6000 48GB PCIe GPUs. Each A6000 GPU offers 768 GB/s memory bandwidth and 1.25 TFLOPs for FP64 operations. Both platforms employ GCC 9.4.0 and CUDA 12.2.

  \textit{Benchmarks.} We employ a wide array of representative QC systems to serve as benchmarks, as listed in Table~\ref{tab:benchmarks}. For correctness validation, we prioritize a diverse mix of systems, encompassing both organic and inorganic molecules. For performance evaluation, we pay particular attention to the computation complexity of these benchmarks. For scalability assessment, the size of the benchmarks is our primary concern. In all evaluations, we utilize the STO-3G basis set, which has relatively lower angular momentum for the sake of simplicity in presentation. It's important to note, however, that \textsc{Matryoshka} is compatible with any basis set.
  
  \begin{table}[]
      \caption{Configuration for \textsc{Matryoshka} benchmarks.}
      \label{tab:benchmarks}
      \begin{adjustbox}{width=0.48\textwidth,center}
      \begin{tabular}{llllll}
      \hline
      \multicolumn{2}{c}{\textbf{Correctness}} & \multicolumn{2}{c}{\textbf{Performance}} & \multicolumn{2}{c}{\textbf{Scalability}} \\
      Name                  & Atoms            & Name                  & Atoms            & Name                  & Atoms            \\
      \hline
      Water                 & 3                & Chinoglin             & 166              & Water Cluster         & 300-11259        \\
      Benzene               & 12               & DNA                   & 566              & GluAla Cluster        & 28-6658          \\
      Water-10              & 30               & Crambin               & 642              &                       &                  \\
      Methanol-7            & 42               & Collagen              & 692              &                       &                  \\
      C60                   & 60               & tRNA                  & 1656             &                       &                  \\
                            &                  & Pepsin                & 2797             &                       &                  \\
      \hline
      \end{tabular}
      \end{adjustbox}
  \end{table}

  \textit{State-of-the-arts.} Due to the majority of work being closed-source or exhibiting significant differences in algorithm implementation, it is challenging to reproduce the results of related work on a one-to-one basis. Despite these obstacles, we have endeavored to collect and reproduce several industry-recognized state-of-the-art works, covering both CPU and GPU for the purpose of quantitative comparison.

  We employ Libint~\cite{CPU-Libint} and PySCF~\cite{CPU-PySCF-1,CPU-PySCF-2,CPU-PySCF-3} as representative examples of CPU-centric designs, which continues to dominate the majority of current work. Libint is one of the most efficient libraries for evaluating molecular integrals. PySCF is a widely utilized quantum chemistry Python package, featuring critical components that are highly optimized in C language. To showcase the emerging trend of GPU-centric work, we opt for QUICK, which stands as the current state-of-the-art for GPU implementations, supported by a series of studies~\cite{GPU-QUICK-1,GPU-QUICK-2,GPU-QUICK-3,GPU-QUICK-4}.
  
\subsection{Correctness Validation}\label{sec:exp-correctness}

  We calculate the total energy for five classic QC systems using \textsc{Matryoshka} as well as other state-of-the-art approaches. In line with the common practice, we set the convergence threshold of electronic density as $10^{-6}$.

  Table~\ref{tab:correctness} shows that the precision of \textsc{Matryoshka} aligns with other state-of-the-art approaches to within $10^{-5}$, meeting the accuracy demands for most QC problems. For C60, a noticeable deviation in the result from QUICK is observed. However, the other approaches including \textsc{Matryoshka} maintain consistency.

\begin{figure}
      \centering
      \includegraphics[width=0.5\textwidth]{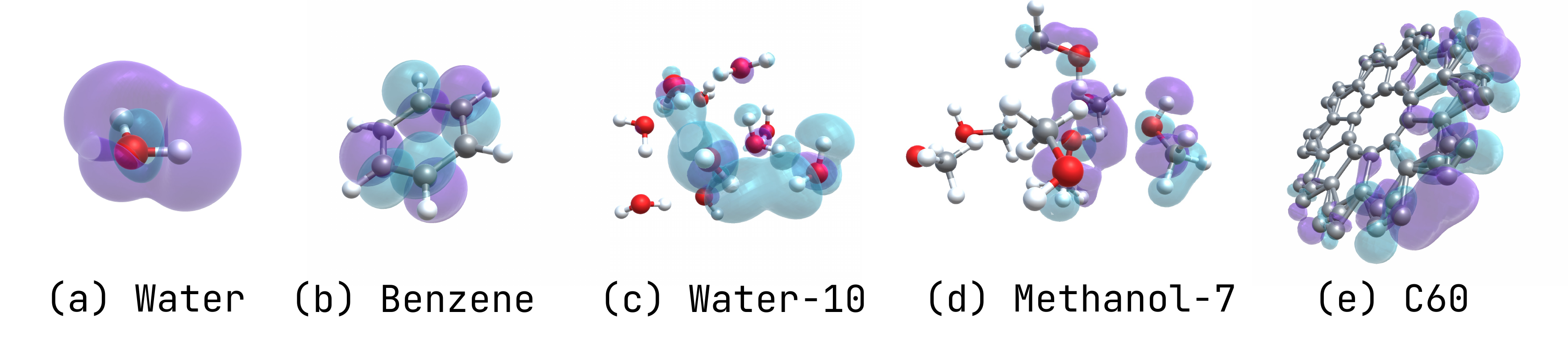}
      \caption{LUMO visualization for five classical QC systems.}
      \label{fig:lumo}
  \end{figure}

  We also visualize the Lowest Unoccupied Molecular Orbital (LUMO) of these QC systems using the results from \textsc{Matryoshka}, as shown in Figure ~\ref{fig:lumo}. These visualizations are in agreement with the experimental data~\cite{ballhausen1965molecular,carey2000advanced}, which reinforces the accuracy of \textsc{Matryoshka}.
  
  \begin{table}
      \caption{Comparison of precision with state-of-the-arts.\tnote{*}}
      \label{tab:correctness}
      \begin{adjustbox}{width=0.5\textwidth,center}
      \begin{threeparttable}
      \begin{tabular}{lllll}
          \hline
          Molecules  & Libint                 & PySCF                  & QUICK                  & \textsc{Matryoshka}    \\ \hline
          Water      & -\underline{\textbf{74.9646977}}   & -\underline{\textbf{74.9646977}}   & -\underline{\textbf{74.9646977}}   & -74.9646977  \\
          Benzene    & -\underline{\textbf{227.8909828}}  & -\underline{\textbf{227.8909828}}  & -\underline{\textbf{227.890982}}7  & -227.8909828  \\
          Water-10   & -\underline{\textbf{749.689879}}0  & -\underline{\textbf{749.689879}}0  & -\underline{\textbf{749.689879}}2  & -749.6898793  \\
          Methanol-7 & -\underline{\textbf{794.773584}}4  & -\underline{\textbf{794.773584}}4  & -\underline{\textbf{794.773584}}3  & -794.7735845  \\
          C60        & -\underline{\textbf{2134.26520}}24 & -\underline{\textbf{2134.26520}}02 & -\underline{\textbf{2134.2}}155554 & -2134.2652033 \\ \hline
      \end{tabular}
      \begin{tablenotes}
          \item [*]The underline highlights the identical digits among the state-of-the-art results.
      \end{tablenotes}
      \end{threeparttable}
      \end{adjustbox}
  \end{table}

\subsection{Breakdown Evaluation}\label{sec:exp-component}

\subsubsection{Performance Breakdown}\label{sec:exp-breakdown}

\begin{figure}
      \centering
      \includegraphics[width=0.5\textwidth]{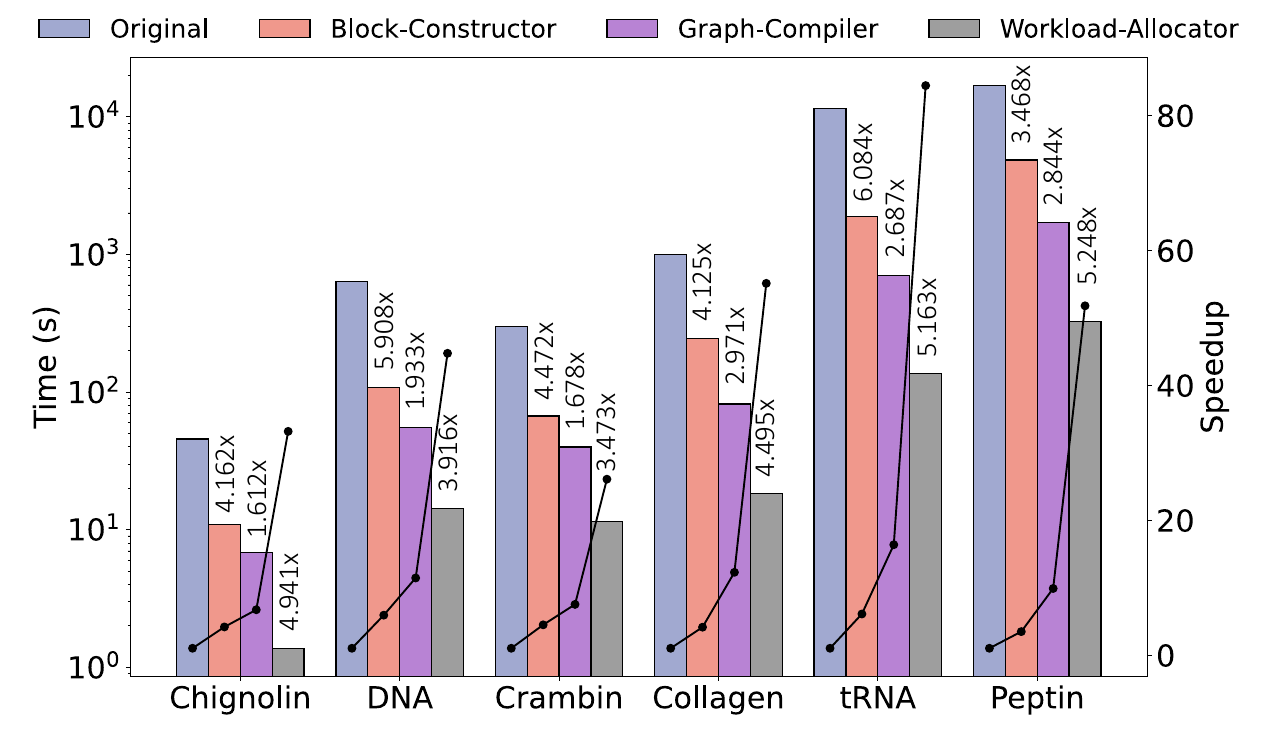}
      \caption{Performance breakdown  of \textsc{Matryoshka}.}
      \label{fig:exp-breakdown}
  \end{figure}

We first investigate how three core components of \textsc{Matryoshka} improve the performance by integrating them progressively. Figure ~\ref{fig:exp-breakdown} shows the performance breakdown of \textsc{Matryoshka} on six representative QC systems with great computational complexity. By clustering ERIs with consistent execution instructions, the Block Constructor mapping the ERI computation onto GPU SIMT architecture efficiently, thus achieving a $4.7\times$ improvement on average. Next, the Graph Compiler deconstructs each ERI into multiple ERI compute tiles and generates the path-clear yet execution-efficient code for them, which brings a $2.3\times$ improvement on average. Finally, the Workload Allocator is integrated which schedules ERI compute tiles to each thread dynamically, which provides a significant $4.5\times$ improvement on average. Here \textsc{Matryoshka} has reached the $33.1\times$, $44.7\times$, $26.1\times$, $55.1\times$, $84.4\times$, $51.8\times$ speedups accumulatively on six representative QC systems.

\subsubsection{Block Constructor}\label{subsec:block-constructor}

  The Block Constructor takes a two-fold approach to complete the construction, systematically organizing the data structures from individual basis functions to pairs, and ultimately to quadruples. This approach significantly reduces the memory cost of ERI computation from the $O(N^4)$ complexity of basis function quadruples to the more manageable $O(N^2)$ of pairs. We count the number of basis function pairs and basis function quadruples in real-world QC systems respectively, as listed in Table ~\ref{tab:exp-block-constructor}. It shows that this nested construction slashes the memory demands by a factor of $10^3$, transforming what was once an infeasible computational task into a feasible one.
  
\begin{figure}
      \centering
      \includegraphics[width=0.40\textwidth]{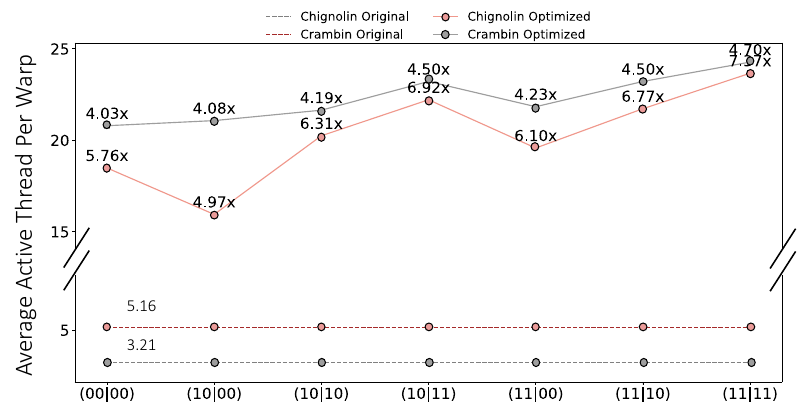}
      \caption{Comparative analysis of Average Active Threads per Warp in two representative QC systems. \textsc{Matryoshka}'s values are classified by ERI classes, while the baseline values are depicted as horizontal lines, indicating the absence of any clustering process.}
      \label{fig:exp-block-constructor}
  \end{figure}

  We choose the metric \textit{average active threads per warp} to study the effectiveness of the Block Constructor in reducing warp divergence, as shown in Figure ~\ref{fig:exp-block-constructor}. A higher number of average active threads per warp indicates less divergence within a single warp. Prior the integration of the Block Constructor, the average active threads per warp were as low as 3.21 and 5.16 for the Chignolin and Crambin respectively, indicating significant warp divergence issues. However, the Block Construct improves this situation markedly by leveraging the Permutation primitive of EPT, achieving increases of up to $7.37\times$ and $4.70\times$ across different ERI classes in the two QC systems.

  \begin{table}
      \caption{The number of basis function pairs and quadruples in six representative QC systems.}
      \label{tab:exp-block-constructor}
      \begin{adjustbox}{width=0.50\textwidth,center}
      \begin{tabular}{lllllll}
          \hline
          QC System & Chignolin & DNA & Crambin & Collagen & tRNA & Peptin \\
          \hline
          Pair & 24.0K & 123.2K & 156.8K & 143.3K & 381.6K & 668.9K \\
          Quadruple & 577.1M & 15.1G & 24.5G & 20.5G & 145.6G & 371.0G \\
          \hline
      \end{tabular}
      \end{adjustbox}
  \end{table}
  
\subsubsection{Graph Compiler}\label{subsec:graph-compiler}

  The Graph Compiler is designed to address two key challenge inherent in a single ERI computation. The first challenge is the severe register spilling due to the complex arithmetic operations involved in ERI computations. To assess the effectiveness of the Graph Compiler in tackling this issue, we measure two specific metrics within the computations of Chignolin and Crambin.

\begin{figure}
    \centering
    \begin{subfigure}{0.45\textwidth}
        \includegraphics[width=\linewidth]{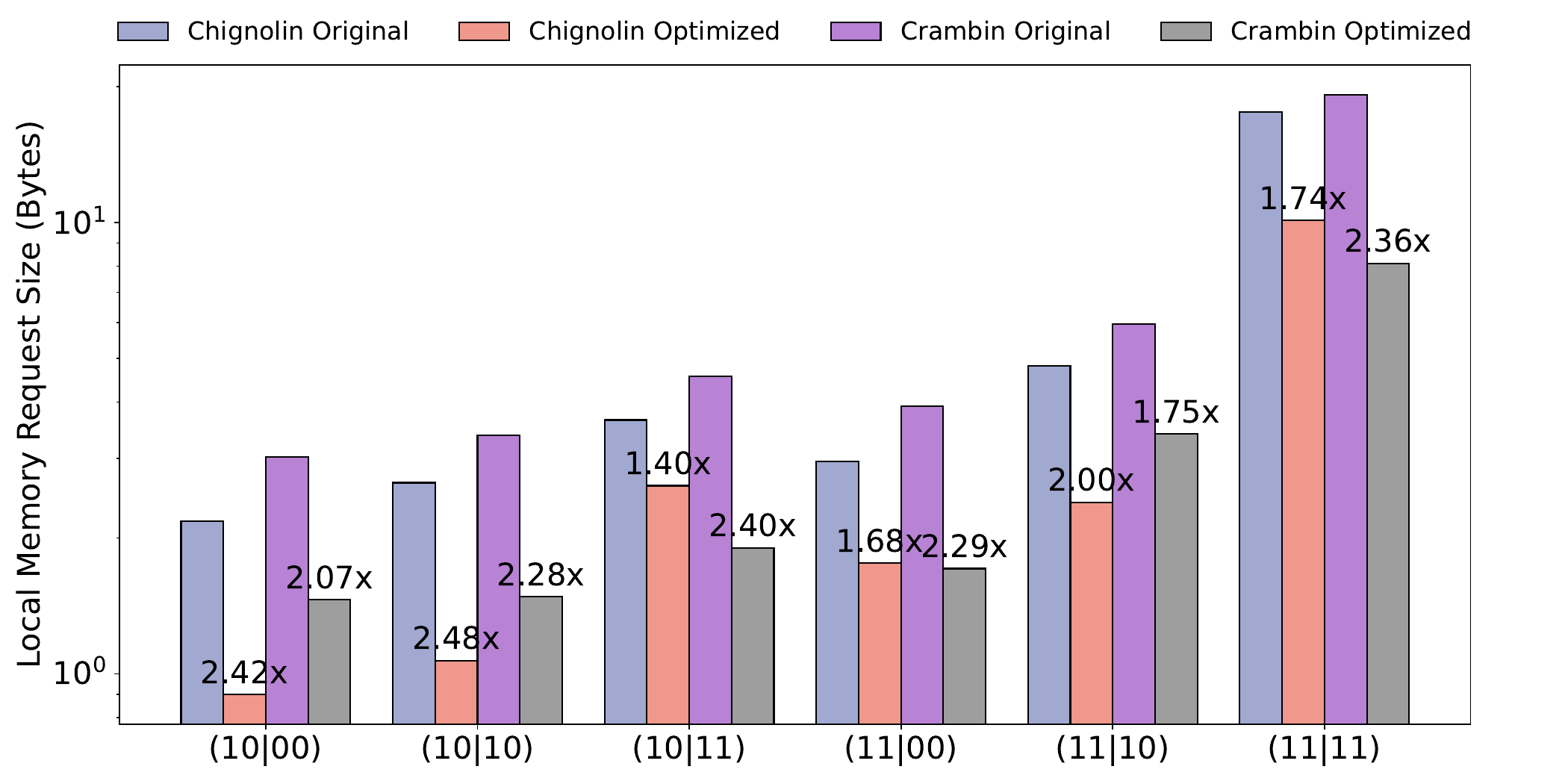}
        \caption{Local memory request.}
        \label{fig:exp-graph-compiler-localmem}
    \end{subfigure}%
    \hfill
    \begin{subfigure}{0.45\textwidth}
        \includegraphics[width=\linewidth]{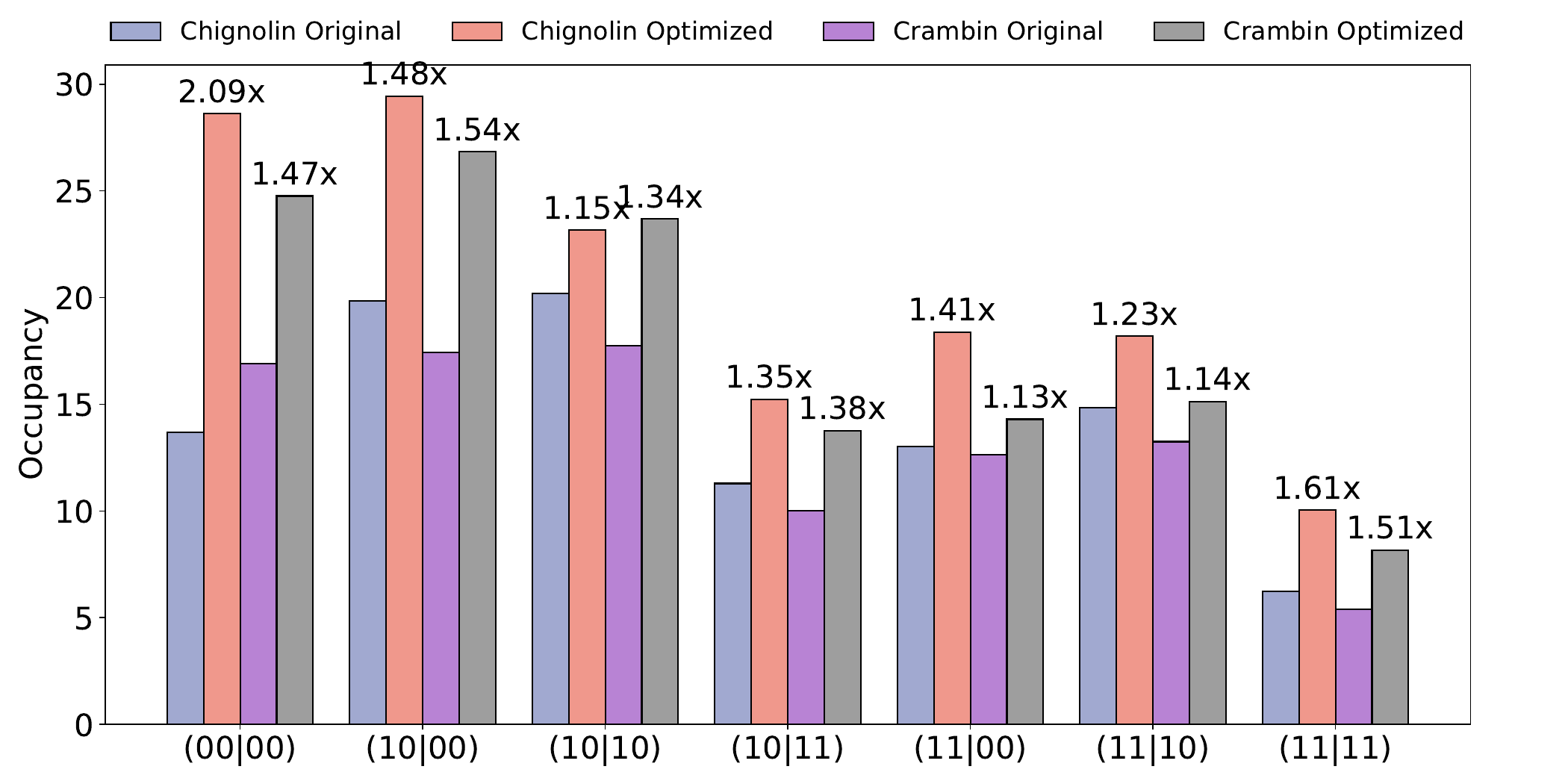}
        \caption{GPU occupancy.}
        \label{fig:exp-graph-compiler-occupancy}
    \end{subfigure}
    \caption{Comparative analysis of local memory request and GPU occupancy on two representative QC systems.}
\end{figure}

  The first metric we examine is the number of local memory requests, as shown in Figure ~\ref{fig:exp-graph-compiler-localmem}. This metric is a direct indicator of the severity of register spilling; when the register file is unable to hold all the data needed for computations, the excess data spills over into the local memory. The result shows that the Graph Compiler, by employing the Deconstruction primitive to partition the complete ERI into several compute tiles, efficiently reduces the local memory requests by up to $2.48\times$ and $2.40\times$ on two QC systems. This substantial decrease in local memory requests implies that the Graph Compiler significantly alleviates the problem of register spilling, thereby enhancing the efficiency of the ERI computation process on these QC systems.

  The second metric we select is the GPU Occupancy, an important metrics that reflects the GPU's ability to hide latencies. Register spilling can negatively impact GPU Occupancy as it increases the number of registers required per thread. Figure ~\ref{fig:exp-graph-compiler-occupancy} shows that the Graph Compiler improves the occupancy of kernels that computes all ERI classes, from $1.23\times$ to $2.09\times$, and from $1.13\times$ to $1.54\times$ on two QC systems respectively.

  The second challenge is the ambiguous computational paths whose computational cost is dynamic. The Graph Compiler uses a greedy path searching algorithm to find an optimized path. Typically, it takes less than 10 seconds for finding the target computational path and generating the corresponding computational kernel. Take the Crambin as an example, the Graph Compiler uses just 2.57 seconds to complete the above tasks within a search space comprising approximately $O(10^5)$ potential computational paths. Notably, this kernel performs $1.42\times$ faster than one derived from a randomly-generated computational path.

\subsubsection{Workload Allocator Evaluation}\label{subsec:workload-allocator}

  Employing the Combination EPT primitive, the Workload Allocator schedules ERI compute tiles at runtime in response to their varying operational intensity. Specially, we measure two metrics to evaluate the effectiveness of the Workload Allocator in Chignolin and Crambin. 
  
\begin{figure}
   \centering
   \begin{subfigure}{0.45\textwidth}
       \includegraphics[width=\linewidth]{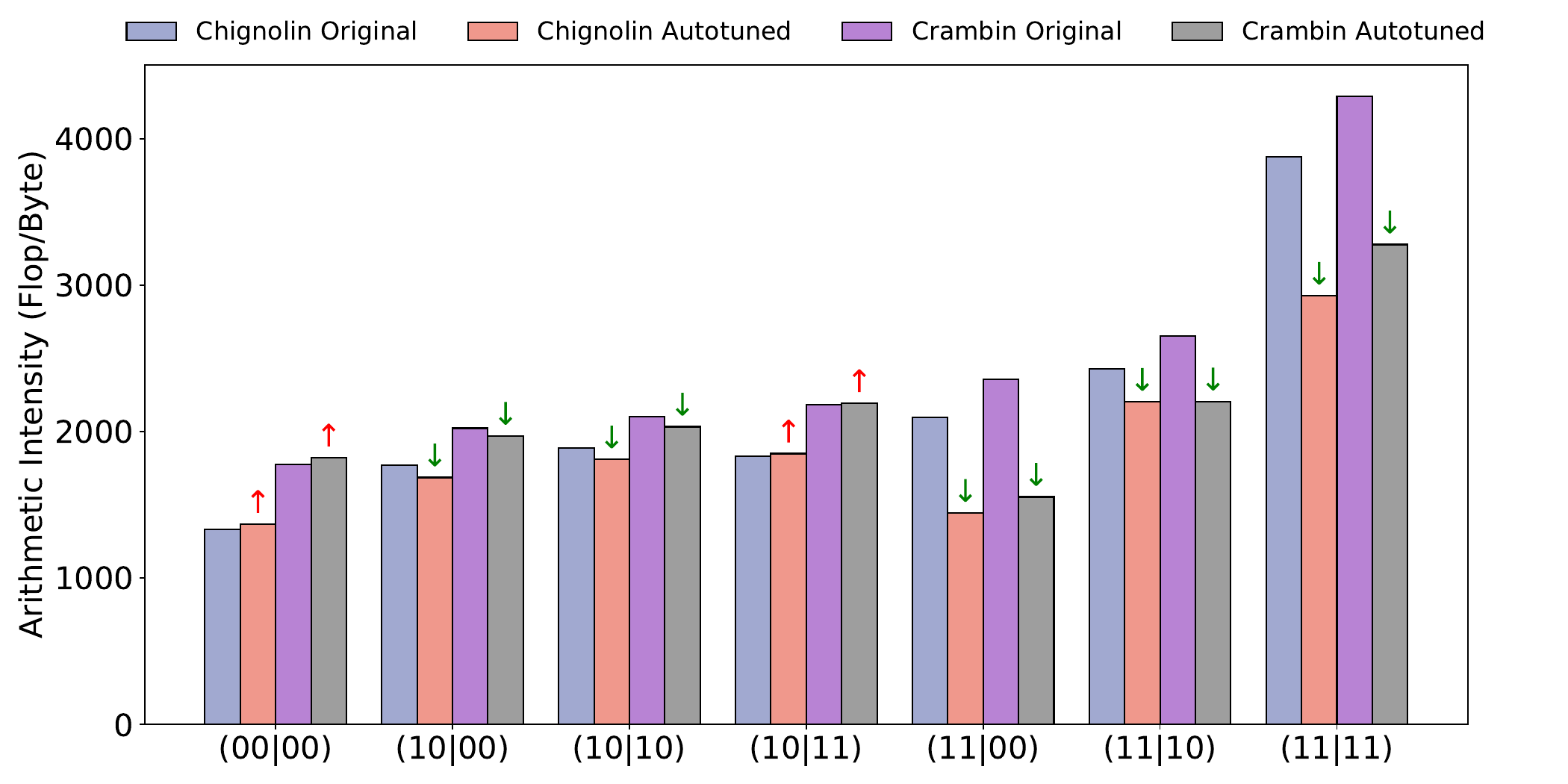}
       \caption{Arithmetic Intensity.}
       \label{fig:exp-workload-allocator-arithmetic-intensity}
   \end{subfigure}

   \begin{subfigure}{0.45\textwidth}
       \includegraphics[width=\linewidth]{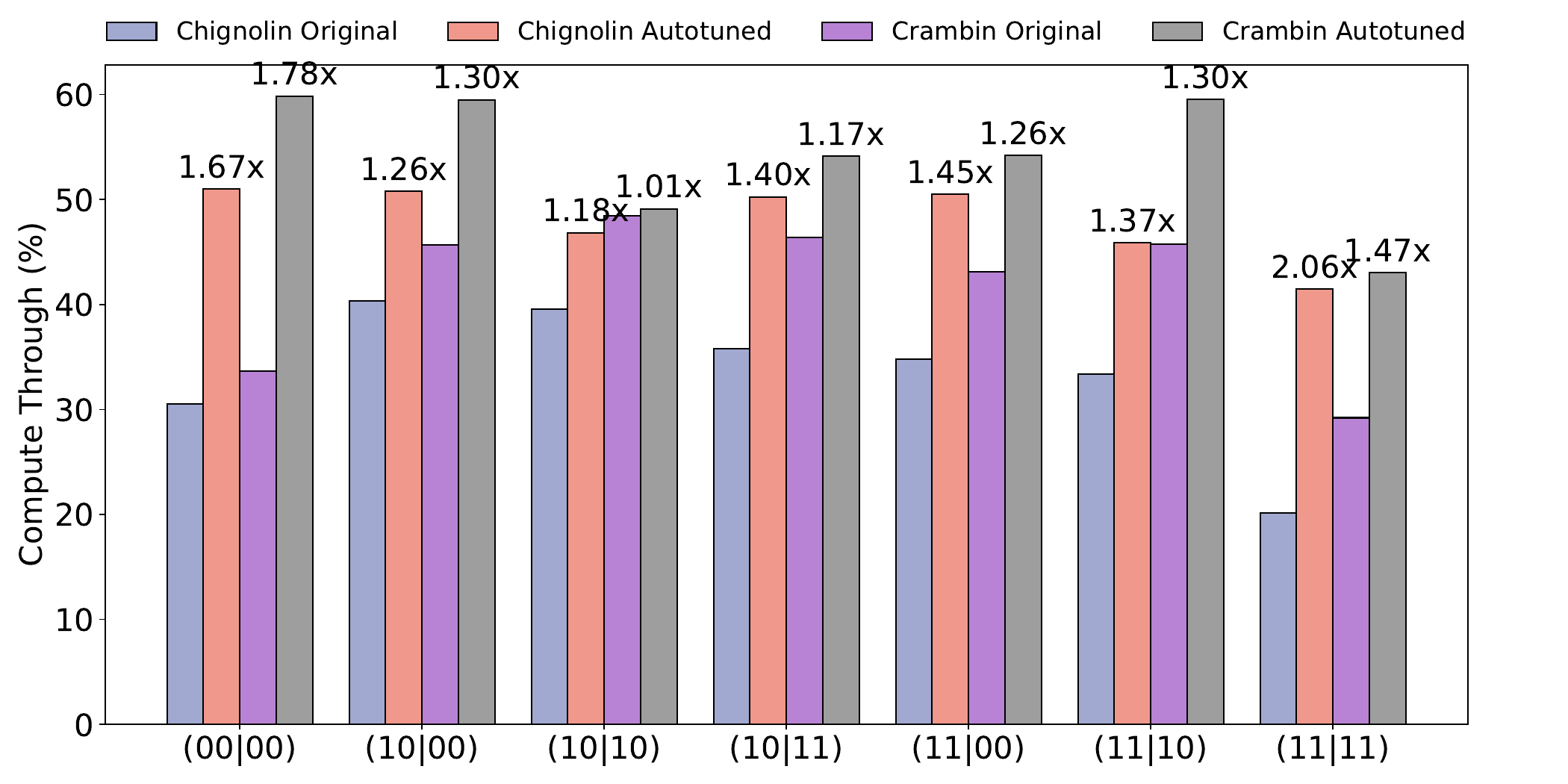}
       \caption{Compute Throughput.}
       \label{fig:exp-workload-allocator-compute-throughput}
   \end{subfigure}
   \caption{Comparative analysis of compute throughput and arithmetic intensity on two representative QC systems.}
\end{figure}

  The first metrics is the arithmetic intensity, which measures the ratio of the FLOP to byte. A higher value indicates greater computational intensity for the kernel. As shown in Figure ~\ref{fig:exp-workload-allocator-arithmetic-intensity},  the Workload Allocator's tuning process adjusts the arithmetic intensity of different ERI classes to address their specific computational bottlenecks.
  
  The other metrics is the compute throughput, which represents the system's computational performance, as shown in Figure ~\ref{fig:exp-workload-allocator-compute-throughput}. The result shows that the Workload Allocator is effective; after tuning, compute throughput across all ERI classes has improved significantly, with increases of up to $2.06\times$ and $1.78\times$, respectively. Additionally, since the auto-tuning algorithm is integrated with the computation process, there is no additional overhead measured.

\subsection{Scalability Evaluation}\label{sec:scalability}

\begin{figure}
      \centering
      \includegraphics[width=0.5\textwidth]{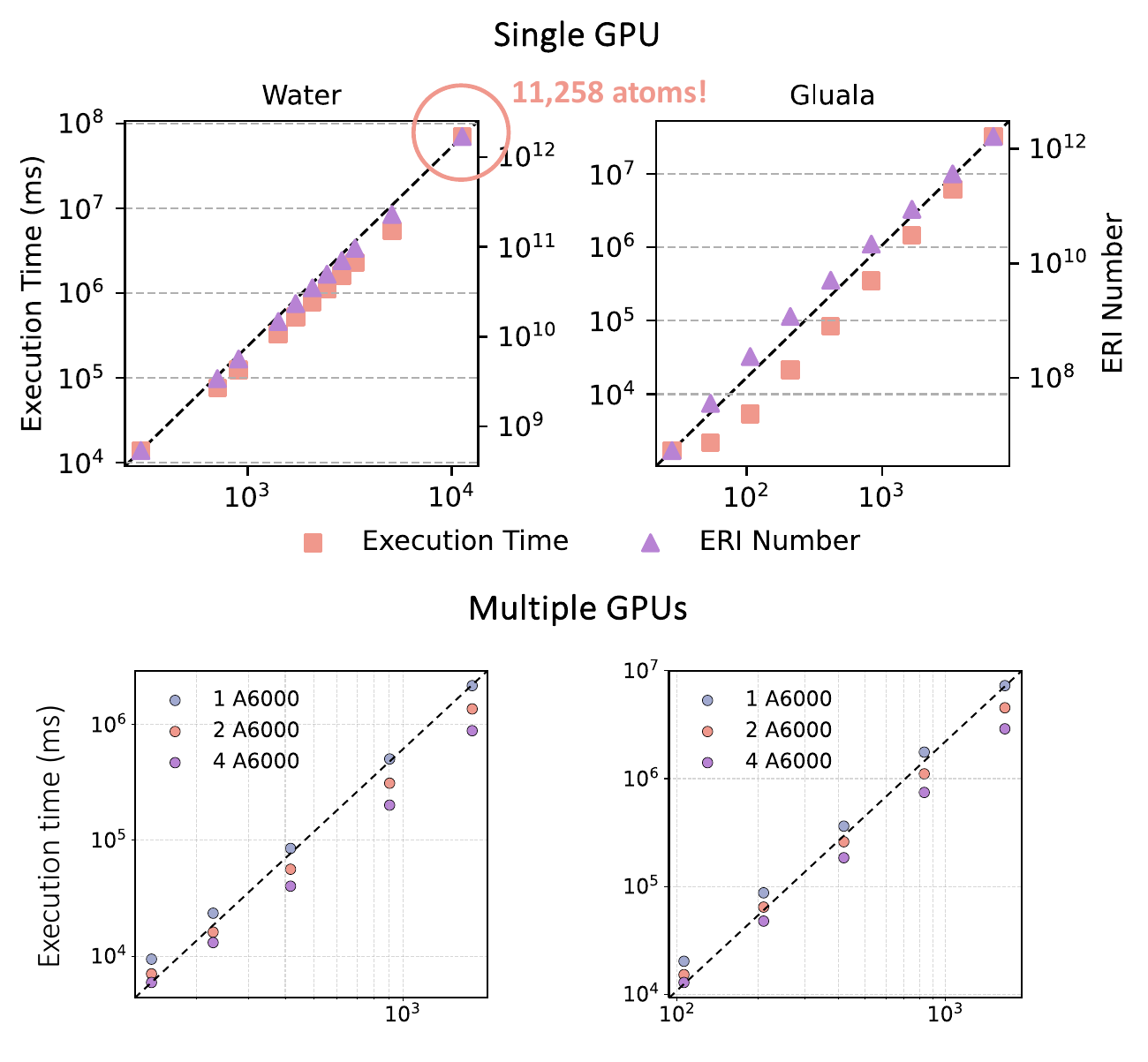}
      \caption{Scalability of \textsc{Matryoshka} on larger QC systems using both a single GPU and multiple GPUs.}
      \label{fig:exp-scalability}
  \end{figure}

  \textit{Single GPU.} We measure the execution time of \textsc{Matryoshka} as it scales with two QC system sizes on single A100 GPU. Figure~\ref{fig:exp-scalability} shows that when the axes are scaled logarithmically, the execution time curve aligns closely with the ERI number curve. This result indicates that both execution time and ERI number grow exponentially with the atom count, while the performance of \textsc{Matryoshka} remains stable despite problem size variations, ensuring consistent efficiency and adaptability across diverse scenarios. It is noteworthy that the largest scale of Water QC system accommodates up to 11,259 atoms and \textsc{Matryoshka} finishes the simulation for 99 iterations in 19.5 hours, which establishes a novel benchmark especially when current studies are predominantly confined to systems with fewer than 1,000 atoms.

  \textit{Multiple GPUs.} We further evaluate the performance of \textsc{Matryoshka} in multi-GPU scenarios with four A6000 GPUs. Figure~\ref{fig:exp-scalability} presents the results of the weak scaling tests on Water and Gluala clusters with increasing molecular scales. With different problem sizes, the speedup of \textsc{Matryoshka} grows approximately proportional to the number of A6000 cards, which demonstrates that the performance of \textsc{Matryoshka} remains consistent regardless of problem size variations, even in multi-GPU scenarios.
  
\subsection{End-to-End Evaluation}\label{sec:end-to-end}
\begin{figure}
      \centering
      \vspace{-8pt}
      \includegraphics[width=0.44\textwidth]{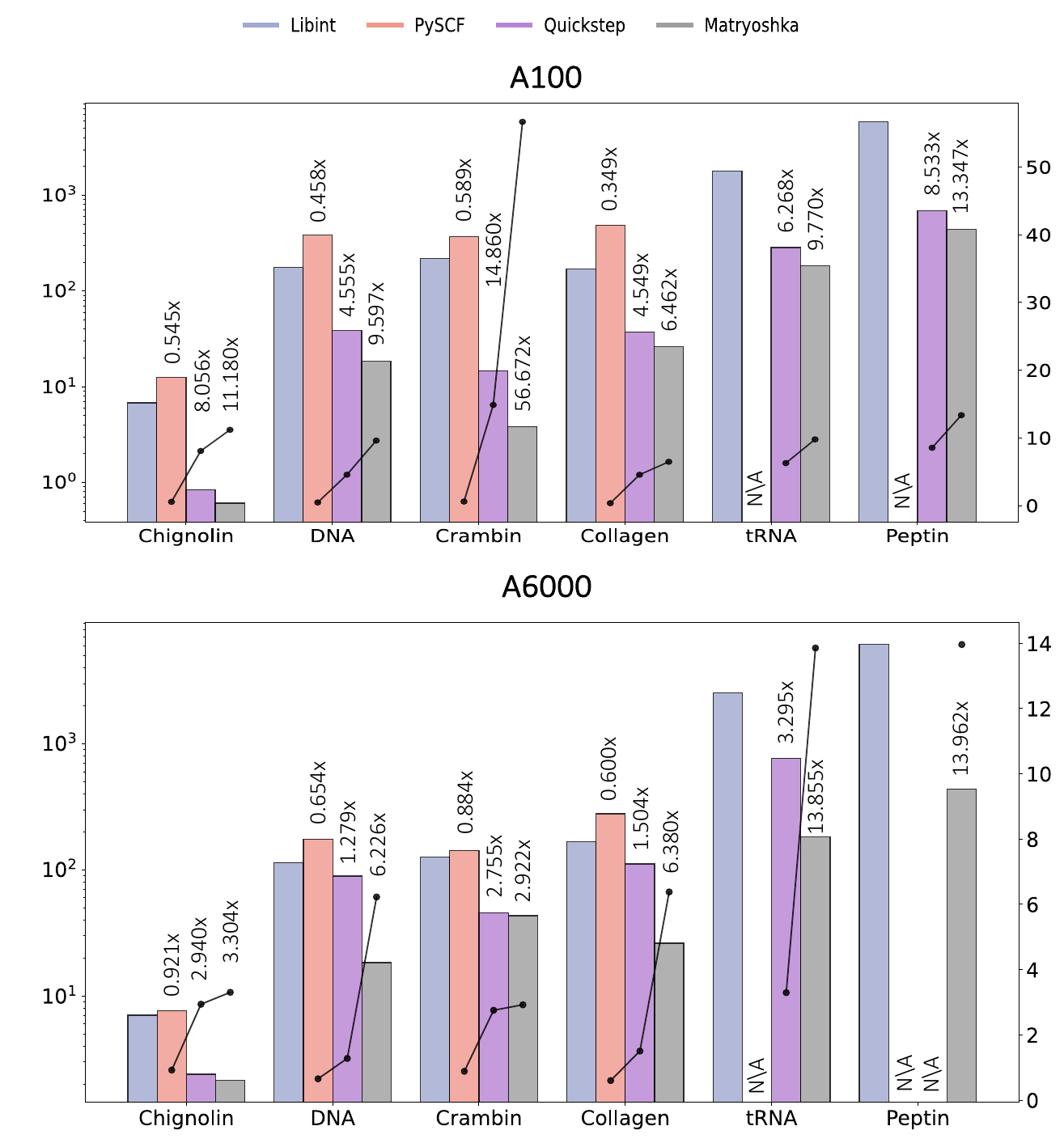}
      \caption{Performance comparison of end-to-end execution time. Speedup is calculated relative to the execution time of Libint.}
      \label{fig:comparison}
  \end{figure}

  Finally, we compare the end-to-end execution time between \textsc{Matryoshka} and other state-of-the-art approaches. To eliminate the impact of different iteration counts, we limit the maximum number of iteration at 99. Figure~\ref{fig:comparison} shows that by capitalizing on EPT, \textsc{Matryoshka} effectively tackles the challenge of dynamic diversity within QC systems and consistently surpasses the previous state-of-the-art approaches across all six representative QC systems on both A100 and A6000 GPUs. The performance of PySCF is slow and insufficient for producing results for large-sized molecules, such as tRNA and Peptin. Libint exhibits superior performance compared to PySCF, owing to more robust multi-thread support. \textsc{Matryoshka} achieves up to $13.35\times$ and $13.86\times$ speedup over Libint on A100 and A6000 GPUs, respectively. Leveraging GPU, QUICK surpasses CPU-based approaches in most benchmarks. However, it performs slower than Libint for the Crambin molecule, emphasizing multi-core CPUs' computational capability. QUICK fails to complete Pepsin calculations due to memory limitations. \textsc{Matryoshka} excels over QUICK across benchmarks, achieving up to $2.11\times$ and $4.82\times$ speedup on A100 and A6000 GPUs.
  
\section{Related Work}

  \textbf{CPU-centric.} Traditional CPU-based approaches~\cite{CPU-Algo-1,CPU-Algo-2,CPU-Algo-3,CPU-Algo-4} primarily focus on minimizing computation and can be divided into two strategies: reducing the number of iterations~\cite{CPU-Algo-1,CPU-Algo-3} and limiting the number of integrals recalculated per iteration~\cite{CPU-Algo-2,CPU-Algo-3}. Although these improvements significantly reduce computation, they do not consider computer hardware, resulting in suboptimal computational efficiency. Recently, several studies~\cite{CPU-GAMESS,CPU-distributed-1,CPU-distributed-2,CPU-distributed-3,CPU-distributed-4,CPU-Load-imbalance-1,CPU-Load-imbalance-2,CPU-Load-imbalance-3,CPU-Load-imbalance-4,CPU-Load-imbalance-5} have concentrated on developing efficient implementations using distributed CPU nodes. Some works aim to design distributed parallel algorithms~\cite{CPU-GAMESS,CPU-distributed-1,CPU-distributed-2,CPU-distributed-3,CPU-distributed-4}, while others address the prevalent load imbalance issues in the field~\cite{CPU-Load-imbalance-1,CPU-Load-imbalance-2,CPU-Load-imbalance-3,CPU-Load-imbalance-4,CPU-Load-imbalance-5}. However, the limitations in computational capability of CPUs hinder their further advancement.
  
  \textbf{GPU-centric.}  The rapid advancements in general-purpose GPU technology have motivated many packages~\cite{related-work-gamess,related-work-nwchem,related-work-gaussian,related-work-cp2k,related-work-libintx-1,related-work-libintx-2,related-work-libintx-3} to transition from traditional CPU to GPU architectures. Additionally, a variety of GPU-specific methods have emerged. Yasudo~\cite{GPU-Yasuda} was among the first to implement GPU technology for the evaluation of ERIs, resulting in substantial speed improvements when compared to CPU-based computations. Following this, Ufimtsev and Martinez~\cite{GPU-Ufimtsev-1,GPU-Ufimtsev-2,GPU-Ufimtsev-3,GPU-Ufimtsev-4} conducted a series of studies that implemented a full Fock build on GPUs, wherein each thread is mapped to a different fundamental integral class. However, their works mainly depended on single-precision arithmetic, leading to considerable computational errors. Asadchev et al.~\cite{GPU-Asadchev} introduced a Rys quadrature ERI method on GPUs that utilized double precision, but overlooked the data transfer costs between CPUs and GPUs. Miao et al.~\cite{GPU-QUICK-1,GPU-QUICK-2,GPU-QUICK-3,GPU-QUICK-4} and Barca et al.~\cite{GPU-Barca} employed the HGP method to reduce the computational expenses associated with the evaluation of ERIs, ultimately achieving remarkable speedups. Despite these advancements, these approaches did not account for the dynamic diversity inherent in QC systems, resulting in under-utilization.

\section{Conclusion}
  This paper presents Matryoshka, a novel elastically-parallel technique for the efficient and accurate execution of quantum chemistry system with dynamic diversity on GPU. Leveraging Elastic Parallelism Transformation, Matryoshka realigns an efficient parallelism with GPU architecture with three key components: The Block Constructor formulates resilient data structures and constructs fine-grained, GPU-efficient compute blocks; the Graph Compiler generates code with clear computational path through an automated compilation process; the Workload Allocator achieves highly efficient parallelism for compute-intensive operations and facilitating automatic fusion with memory-intensive operations online.  


\bibliographystyle{ACM-Reference-Format}
\bibliography{reference}


\begin{thebibliography}{65}


\ifx \showCODEN    \undefined \def \showCODEN     #1{\unskip}     \fi
\ifx \showDOI      \undefined \def \showDOI       #1{#1}\fi
\ifx \showISBNx    \undefined \def \showISBNx     #1{\unskip}     \fi
\ifx \showISBNxiii \undefined \def \showISBNxiii  #1{\unskip}     \fi
\ifx \showISSN     \undefined \def \showISSN      #1{\unskip}     \fi
\ifx \showLCCN     \undefined \def \showLCCN      #1{\unskip}     \fi
\ifx \shownote     \undefined \def \shownote      #1{#1}          \fi
\ifx \showarticletitle \undefined \def \showarticletitle #1{#1}   \fi
\ifx \showURL      \undefined \def \showURL       {\relax}        \fi
\providecommand\bibfield[2]{#2}
\providecommand\bibinfo[2]{#2}
\providecommand\natexlab[1]{#1}
\providecommand\showeprint[2][]{arXiv:#2}

\bibitem[Alexeev et~al\mbox{.}(2002)]%
        {CPU-distributed-2}
\bibfield{author}{\bibinfo{person}{Yuri Alexeev}, \bibinfo{person}{Ricky~A. Kendall}, {and} \bibinfo{person}{Mark~S. Gordon}.} \bibinfo{year}{2002}\natexlab{}.
\newblock \showarticletitle{The distributed data SCF}.
\newblock \bibinfo{journal}{\emph{Computer Physics Communications}} \bibinfo{volume}{143}, \bibinfo{number}{1} (\bibinfo{year}{2002}), \bibinfo{pages}{69--82}.
\newblock
\showISSN{0010-4655}
\urldef\tempurl%
\url{https://doi.org/10.1016/S0010-4655(01)00439-8}
\showDOI{\tempurl}


\bibitem[Alexeev et~al\mbox{.}(2014)]%
        {CPU-Load-imbalance-3}
\bibfield{author}{\bibinfo{person}{Yuri Alexeev}, \bibinfo{person}{Sheri Mickelson}, \bibinfo{person}{Sven Leyffer}, \bibinfo{person}{Robert Jacob}, {and} \bibinfo{person}{Anthony Craig}.} \bibinfo{year}{2014}\natexlab{}.
\newblock \showarticletitle{The Heuristic Static Load-Balancing Algorithm Applied to the Community Earth System Model}. In \bibinfo{booktitle}{\emph{2014 IEEE International Parallel \& Distributed Processing Symposium Workshops}}. \bibinfo{pages}{1581--1590}.
\newblock
\urldef\tempurl%
\url{https://doi.org/10.1109/IPDPSW.2014.177}
\showDOI{\tempurl}


\bibitem[Alexeev et~al\mbox{.}(2007)]%
        {CPU-distributed-3}
\bibfield{author}{\bibinfo{person}{Yuri Alexeev}, \bibinfo{person}{Michael~W. Schmidt}, \bibinfo{person}{Theresa~L. Windus}, {and} \bibinfo{person}{Mark~S. Gordon}.} \bibinfo{year}{2007}\natexlab{}.
\newblock \showarticletitle{A parallel distributed data CPHF algorithm for analytic Hessians}.
\newblock \bibinfo{journal}{\emph{Journal of Computational Chemistry}} \bibinfo{volume}{28}, \bibinfo{number}{10} (\bibinfo{year}{2007}), \bibinfo{pages}{1685--1694}.
\newblock
\urldef\tempurl%
\url{https://doi.org/10.1002/jcc.20633}
\showDOI{\tempurl}
\showeprint{https://onlinelibrary.wiley.com/doi/pdf/10.1002/jcc.20633}


\bibitem[Almlöf et~al\mbox{.}(1982)]%
        {CPU-Algo-1}
\bibfield{author}{\bibinfo{person}{J. Almlöf}, \bibinfo{person}{K. Faegri~Jr.}, {and} \bibinfo{person}{K. Korsell}.} \bibinfo{year}{1982}\natexlab{}.
\newblock \showarticletitle{Principles for a direct SCF approach to LICAO–MOab-initio calculations}.
\newblock \bibinfo{journal}{\emph{Journal of Computational Chemistry}} \bibinfo{volume}{3}, \bibinfo{number}{3} (\bibinfo{year}{1982}), \bibinfo{pages}{385--399}.
\newblock
\urldef\tempurl%
\url{https://doi.org/10.1002/jcc.540030314}
\showDOI{\tempurl}
\showeprint{https://onlinelibrary.wiley.com/doi/pdf/10.1002/jcc.540030314}


\bibitem[Ando et~al\mbox{.}(2021)]%
        {cpu-covid}
\bibfield{author}{\bibinfo{person}{Kazuto Ando}, \bibinfo{person}{Rahul Bale}, \bibinfo{person}{ChungGang Li}, \bibinfo{person}{Satoshi Matsuoka}, \bibinfo{person}{Keiji Onishi}, {and} \bibinfo{person}{Makoto Tsubokura}.} \bibinfo{year}{2021}\natexlab{}.
\newblock \showarticletitle{Digital transformation of droplet/aerosol infection risk assessment realized on" Fugaku" for the fight against COVID-19}.
\newblock \bibinfo{journal}{\emph{arXiv preprint arXiv:2110.09769}} (\bibinfo{year}{2021}).
\newblock


\bibitem[Asadchev et~al\mbox{.}(2010)]%
        {GPU-Asadchev}
\bibfield{author}{\bibinfo{person}{Andrey Asadchev}, \bibinfo{person}{Veerendra Allada}, \bibinfo{person}{Jacob Felder}, \bibinfo{person}{Brett~M. Bode}, \bibinfo{person}{Mark~S. Gordon}, {and} \bibinfo{person}{Theresa~L. Windus}.} \bibinfo{year}{2010}\natexlab{}.
\newblock \showarticletitle{Uncontracted Rys Quadrature Implementation of up to G Functions on Graphical Processing Units}.
\newblock \bibinfo{journal}{\emph{Journal of Chemical Theory and Computation}} \bibinfo{volume}{6}, \bibinfo{number}{3} (\bibinfo{year}{2010}), \bibinfo{pages}{696--704}.
\newblock
\urldef\tempurl%
\url{https://doi.org/10.1021/ct9005079}
\showDOI{\tempurl}
\showeprint{https://doi.org/10.1021/ct9005079}
\newblock
\shownote{PMID: 26613300}.


\bibitem[Asadchev and Valeev(2023a)]%
        {related-work-libintx-1}
\bibfield{author}{\bibinfo{person}{Andrey Asadchev} {and} \bibinfo{person}{Edward~F Valeev}.} \bibinfo{year}{2023}\natexlab{a}.
\newblock \showarticletitle{High-performance evaluation of high angular momentum 4-center Gaussian integrals on modern accelerated processors}.
\newblock \bibinfo{journal}{\emph{The Journal of Physical Chemistry A}} \bibinfo{volume}{127}, \bibinfo{number}{51} (\bibinfo{year}{2023}), \bibinfo{pages}{10889--10895}.
\newblock


\bibitem[Asadchev and Valeev(2023b)]%
        {related-work-libintx-2}
\bibfield{author}{\bibinfo{person}{Andrey Asadchev} {and} \bibinfo{person}{Edward~F Valeev}.} \bibinfo{year}{2023}\natexlab{b}.
\newblock \showarticletitle{Memory-efficient recursive evaluation of 3-center Gaussian integrals}.
\newblock \bibinfo{journal}{\emph{Journal of Chemical Theory and Computation}} \bibinfo{volume}{19}, \bibinfo{number}{6} (\bibinfo{year}{2023}), \bibinfo{pages}{1698--1710}.
\newblock


\bibitem[Asadchev and Valeev(2024)]%
        {related-work-libintx-3}
\bibfield{author}{\bibinfo{person}{Andrey Asadchev} {and} \bibinfo{person}{Edward~F Valeev}.} \bibinfo{year}{2024}\natexlab{}.
\newblock \showarticletitle{3-center and 4-center 2-particle Gaussian AO integrals on modern accelerated processors}.
\newblock \bibinfo{journal}{\emph{arXiv preprint arXiv:2405.01834}} (\bibinfo{year}{2024}).
\newblock


\bibitem[Ballhausen and Gray(1965)]%
        {ballhausen1965molecular}
\bibfield{author}{\bibinfo{person}{C.J. Ballhausen} {and} \bibinfo{person}{H.B. Gray}.} \bibinfo{year}{1965}\natexlab{}.
\newblock \bibinfo{booktitle}{\emph{Molecular Orbital Theory: An Introductory Lecture Note and Reprint Volume}}.
\newblock \bibinfo{publisher}{W.A. Benjamin}.
\newblock
\showISBNx{9780805304510}
\showLCCN{lc65012062}
\urldef\tempurl%
\url{https://books.google.com/books?id=HYw-AAAAIAAJ}
\showURL{%
\tempurl}


\bibitem[Barca et~al\mbox{.}(2020)]%
        {GPU-Barca}
\bibfield{author}{\bibinfo{person}{Giuseppe M.~J. Barca}, \bibinfo{person}{Jorge~L. Galvez-Vallejo}, \bibinfo{person}{David~L. Poole}, \bibinfo{person}{Alistair~P. Rendell}, {and} \bibinfo{person}{Mark~S. Gordon}.} \bibinfo{year}{2020}\natexlab{}.
\newblock \showarticletitle{High-Performance, Graphics Processing Unit-Accelerated Fock Build Algorithm}.
\newblock \bibinfo{journal}{\emph{Journal of Chemical Theory and Computation}} \bibinfo{volume}{16}, \bibinfo{number}{12} (\bibinfo{year}{2020}), \bibinfo{pages}{7232--7238}.
\newblock
\urldef\tempurl%
\url{https://doi.org/10.1021/acs.jctc.0c00768}
\showDOI{\tempurl}
\showeprint{https://doi.org/10.1021/acs.jctc.0c00768}
\newblock
\shownote{PMID: 33206515}.


\bibitem[Cances et~al\mbox{.}(2003)]%
        {QC-equation}
\bibfield{author}{\bibinfo{person}{Eric Cances}, \bibinfo{person}{Mireille Defranceschi}, \bibinfo{person}{Werner Kutzelnigg}, \bibinfo{person}{Claude Le~Bris}, {and} \bibinfo{person}{Yvon Maday}.} \bibinfo{year}{2003}\natexlab{}.
\newblock \showarticletitle{Computational quantum chemistry: a primer}.
\newblock \bibinfo{journal}{\emph{Handbook of numerical analysis}}  \bibinfo{volume}{10} (\bibinfo{year}{2003}), \bibinfo{pages}{3--270}.
\newblock


\bibitem[Cao et~al\mbox{.}(2019)]%
        {QC-review-1}
\bibfield{author}{\bibinfo{person}{Yudong Cao}, \bibinfo{person}{Jonathan Romero}, \bibinfo{person}{Jonathan~P Olson}, \bibinfo{person}{Matthias Degroote}, \bibinfo{person}{Peter~D Johnson}, \bibinfo{person}{M{\'a}ria Kieferov{\'a}}, \bibinfo{person}{Ian~D Kivlichan}, \bibinfo{person}{Tim Menke}, \bibinfo{person}{Borja Peropadre}, \bibinfo{person}{Nicolas~PD Sawaya}, {et~al\mbox{.}}} \bibinfo{year}{2019}\natexlab{}.
\newblock \showarticletitle{Quantum chemistry in the age of quantum computing}.
\newblock \bibinfo{journal}{\emph{Chemical reviews}} \bibinfo{volume}{119}, \bibinfo{number}{19} (\bibinfo{year}{2019}), \bibinfo{pages}{10856--10915}.
\newblock


\bibitem[Carey and Sundberg(2000)]%
        {carey2000advanced}
\bibfield{author}{\bibinfo{person}{F.A. Carey} {and} \bibinfo{person}{R.J. Sundberg}.} \bibinfo{year}{2000}\natexlab{}.
\newblock \bibinfo{booktitle}{\emph{Advanced Organic Chemistry: Part A: Structure and Mechanisms}}.
\newblock \bibinfo{publisher}{Kluwer Academic/Plenum Pub.}
\newblock
\showISBNx{9780306462436}
\showLCCN{00027456}
\urldef\tempurl%
\url{https://books.google.com/books?id=z3jXKOYuqQAC}
\showURL{%
\tempurl}


\bibitem[Carloni et~al\mbox{.}(2002)]%
        {QC-pharmaceuticals-1}
\bibfield{author}{\bibinfo{person}{Paolo Carloni}, \bibinfo{person}{Ursula Rothlisberger}, {and} \bibinfo{person}{Michele Parrinello}.} \bibinfo{year}{2002}\natexlab{}.
\newblock \showarticletitle{The Role and Perspective of Ab Initio Molecular Dynamics in the Study of Biological Systems}.
\newblock \bibinfo{journal}{\emph{Accounts of Chemical Research}} \bibinfo{volume}{35}, \bibinfo{number}{6} (\bibinfo{year}{2002}), \bibinfo{pages}{455--464}.
\newblock
\urldef\tempurl%
\url{https://doi.org/10.1021/ar010018u}
\showDOI{\tempurl}
\showeprint{https://doi.org/10.1021/ar010018u}
\newblock
\shownote{PMID: 12069631}.


\bibitem[Chow et~al\mbox{.}(2015)]%
        {CPU-Load-imbalance-2}
\bibfield{author}{\bibinfo{person}{Edmond Chow}, \bibinfo{person}{Xing Liu}, \bibinfo{person}{Sanchit Misra}, \bibinfo{person}{Marat Dukhan}, \bibinfo{person}{Mikhail Smelyanskiy}, \bibinfo{person}{Jeff Hammond}, \bibinfo{person}{Yunfei Du}, \bibinfo{person}{Xiang-Ke Liao}, {and} \bibinfo{person}{Pradeep Dubey}.} \bibinfo{year}{2015}\natexlab{}.
\newblock \showarticletitle{Scaling up Hartree-Fock calculations on Tianhe-2}.
\newblock \bibinfo{journal}{\emph{International Journal of High Performance Computing Applications}}  \bibinfo{volume}{30} (\bibinfo{date}{07} \bibinfo{year}{2015}).
\newblock
\urldef\tempurl%
\url{https://doi.org/10.1177/1094342015592960}
\showDOI{\tempurl}


\bibitem[Cremer and Gauss(1986)]%
        {CPU-Algo-3}
\bibfield{author}{\bibinfo{person}{Dieter Cremer} {and} \bibinfo{person}{JüRgen Gauss}.} \bibinfo{year}{1986}\natexlab{}.
\newblock \showarticletitle{An unconventional scf method for calculations on large molecules}.
\newblock \bibinfo{journal}{\emph{Journal of Computational Chemistry}} \bibinfo{volume}{7}, \bibinfo{number}{3} (\bibinfo{year}{1986}), \bibinfo{pages}{274--282}.
\newblock
\urldef\tempurl%
\url{https://doi.org/10.1002/jcc.540070305}
\showDOI{\tempurl}
\showeprint{https://onlinelibrary.wiley.com/doi/pdf/10.1002/jcc.540070305}


\bibitem[Das et~al\mbox{.}(2023)]%
        {sc-gb23}
\bibfield{author}{\bibinfo{person}{Sambit Das}, \bibinfo{person}{Bikash Kanungo}, \bibinfo{person}{Vishal Subramanian}, \bibinfo{person}{Gourab Panigrahi}, \bibinfo{person}{Phani Motamarri}, \bibinfo{person}{David Rogers}, \bibinfo{person}{Paul Zimmerman}, {and} \bibinfo{person}{Vikram Gavini}.} \bibinfo{year}{2023}\natexlab{}.
\newblock \showarticletitle{Large-Scale Materials Modeling at Quantum Accuracy: Ab Initio Simulations of Quasicrystals and Interacting Extended Defects in Metallic Alloys} \emph{(\bibinfo{series}{SC '23})}. \bibinfo{publisher}{Association for Computing Machinery}, \bibinfo{address}{New York, NY, USA}, Article \bibinfo{articleno}{1}, \bibinfo{numpages}{12}~pages.
\newblock
\showISBNx{9798400701092}
\urldef\tempurl%
\url{https://doi.org/10.1145/3581784.3627037}
\showDOI{\tempurl}


\bibitem[Datta et~al\mbox{.}(2008)]%
        {cpu-stencil-1}
\bibfield{author}{\bibinfo{person}{Kaushik Datta}, \bibinfo{person}{Mark Murphy}, \bibinfo{person}{Vasily Volkov}, \bibinfo{person}{Samuel Williams}, \bibinfo{person}{Jonathan Carter}, \bibinfo{person}{Leonid Oliker}, \bibinfo{person}{David Patterson}, \bibinfo{person}{John Shalf}, {and} \bibinfo{person}{Katherine Yelick}.} \bibinfo{year}{2008}\natexlab{}.
\newblock \showarticletitle{Stencil computation optimization and auto-tuning on state-of-the-art multicore architectures}. In \bibinfo{booktitle}{\emph{SC'08: Proceedings of the 2008 ACM/IEEE conference on Supercomputing}}. IEEE, \bibinfo{pages}{1--12}.
\newblock


\bibitem[Devlin et~al\mbox{.}(2018)]%
        {ai-bert}
\bibfield{author}{\bibinfo{person}{Jacob Devlin}, \bibinfo{person}{Ming-Wei Chang}, \bibinfo{person}{Kenton Lee}, {and} \bibinfo{person}{Kristina Toutanova}.} \bibinfo{year}{2018}\natexlab{}.
\newblock \showarticletitle{Bert: Pre-training of deep bidirectional transformers for language understanding}.
\newblock \bibinfo{journal}{\emph{arXiv preprint arXiv:1810.04805}} (\bibinfo{year}{2018}).
\newblock


\bibitem[Fedeli et~al\mbox{.}(2022)]%
        {sc-gb22}
\bibfield{author}{\bibinfo{person}{L. Fedeli}, \bibinfo{person}{A. Huebl}, \bibinfo{person}{F. Boillod-Cerneux}, \bibinfo{person}{T. Clark}, \bibinfo{person}{K. Gott}, \bibinfo{person}{C. Hillairet}, \bibinfo{person}{S. Jaure}, \bibinfo{person}{A. Leblanc}, \bibinfo{person}{R. Lehe}, \bibinfo{person}{A. Myers}, \bibinfo{person}{C. Piechurski}, \bibinfo{person}{M. Sato}, \bibinfo{person}{N. Zaim}, \bibinfo{person}{W. Zhang}, \bibinfo{person}{J. Vay}, {and} \bibinfo{person}{H. Vincenti}.} \bibinfo{year}{2022}\natexlab{}.
\newblock \showarticletitle{Pushing the Frontier in the Design of Laser-Based Electron Accelerators with Groundbreaking Mesh-Refined Particle-In-Cell Simulations on Exascale-Class Supercomputers}. In \bibinfo{booktitle}{\emph{SC22: International Conference for High Performance Computing, Networking, Storage and Analysis}}. \bibinfo{publisher}{IEEE Computer Society}, \bibinfo{address}{Los Alamitos, CA, USA}, \bibinfo{pages}{1--12}.
\newblock
\urldef\tempurl%
\url{https://doi.org/10.1109/SC41404.2022.00008}
\showDOI{\tempurl}


\bibitem[Fletcher et~al\mbox{.}(2000)]%
        {CPU-distributed-4}
\bibfield{author}{\bibinfo{person}{Graham~D. Fletcher}, \bibinfo{person}{Michael~W. Schmidt}, \bibinfo{person}{Brett~M. Bode}, {and} \bibinfo{person}{Mark~S. Gordon}.} \bibinfo{year}{2000}\natexlab{}.
\newblock \showarticletitle{The Distributed Data Interface in GAMESS}.
\newblock \bibinfo{journal}{\emph{Computer Physics Communications}} \bibinfo{volume}{128}, \bibinfo{number}{1} (\bibinfo{year}{2000}), \bibinfo{pages}{190--200}.
\newblock
\showISSN{0010-4655}
\urldef\tempurl%
\url{https://doi.org/10.1016/S0010-4655(00)00073-4}
\showDOI{\tempurl}


\bibitem[Frisch et~al\mbox{.}(2016)]%
        {related-work-gaussian}
\bibfield{author}{\bibinfo{person}{M.~J. Frisch}, \bibinfo{person}{G.~W. Trucks}, \bibinfo{person}{H.~B. Schlegel}, \bibinfo{person}{G.~E. Scuseria}, \bibinfo{person}{M.~A. Robb}, \bibinfo{person}{J.~R. Cheeseman}, \bibinfo{person}{G. Scalmani}, \bibinfo{person}{V. Barone}, \bibinfo{person}{G.~A. Petersson}, \bibinfo{person}{H. Nakatsuji}, \bibinfo{person}{X. Li}, \bibinfo{person}{M. Caricato}, \bibinfo{person}{A.~V. Marenich}, \bibinfo{person}{J. Bloino}, \bibinfo{person}{B.~G. Janesko}, \bibinfo{person}{R. Gomperts}, \bibinfo{person}{B. Mennucci}, \bibinfo{person}{H.~P. Hratchian}, \bibinfo{person}{J.~V. Ortiz}, \bibinfo{person}{A.~F. Izmaylov}, \bibinfo{person}{J.~L. Sonnenberg}, \bibinfo{person}{D. Williams-Young}, \bibinfo{person}{F. Ding}, \bibinfo{person}{F. Lipparini}, \bibinfo{person}{F. Egidi}, \bibinfo{person}{J. Goings}, \bibinfo{person}{B. Peng}, \bibinfo{person}{A. Petrone}, \bibinfo{person}{T. Henderson}, \bibinfo{person}{D. Ranasinghe}, \bibinfo{person}{V.~G. Zakrzewski}, \bibinfo{person}{J.
  Gao}, \bibinfo{person}{N. Rega}, \bibinfo{person}{G. Zheng}, \bibinfo{person}{W. Liang}, \bibinfo{person}{M. Hada}, \bibinfo{person}{M. Ehara}, \bibinfo{person}{K. Toyota}, \bibinfo{person}{R. Fukuda}, \bibinfo{person}{J. Hasegawa}, \bibinfo{person}{M. Ishida}, \bibinfo{person}{T. Nakajima}, \bibinfo{person}{Y. Honda}, \bibinfo{person}{O. Kitao}, \bibinfo{person}{H. Nakai}, \bibinfo{person}{T. Vreven}, \bibinfo{person}{K. Throssell}, \bibinfo{person}{J.~A. Montgomery, {Jr.}}, \bibinfo{person}{J.~E. Peralta}, \bibinfo{person}{F. Ogliaro}, \bibinfo{person}{M.~J. Bearpark}, \bibinfo{person}{J.~J. Heyd}, \bibinfo{person}{E.~N. Brothers}, \bibinfo{person}{K.~N. Kudin}, \bibinfo{person}{V.~N. Staroverov}, \bibinfo{person}{T.~A. Keith}, \bibinfo{person}{R. Kobayashi}, \bibinfo{person}{J. Normand}, \bibinfo{person}{K. Raghavachari}, \bibinfo{person}{A.~P. Rendell}, \bibinfo{person}{J.~C. Burant}, \bibinfo{person}{S.~S. Iyengar}, \bibinfo{person}{J. Tomasi}, \bibinfo{person}{M. Cossi}, \bibinfo{person}{J.~M.
  Millam}, \bibinfo{person}{M. Klene}, \bibinfo{person}{C. Adamo}, \bibinfo{person}{R. Cammi}, \bibinfo{person}{J.~W. Ochterski}, \bibinfo{person}{R.~L. Martin}, \bibinfo{person}{K. Morokuma}, \bibinfo{person}{O. Farkas}, \bibinfo{person}{J.~B. Foresman}, {and} \bibinfo{person}{D.~J. Fox}.} \bibinfo{year}{2016}\natexlab{}.
\newblock \bibinfo{title}{Gaussian˜16 {R}evision {C}.01}.
\newblock
\newblock
\newblock
\shownote{Gaussian Inc. Wallingford CT}.


\bibitem[Giani and Eldredge(2021)]%
        {QC-energy-production}
\bibfield{author}{\bibinfo{person}{Annarita Giani} {and} \bibinfo{person}{Zachary Eldredge}.} \bibinfo{year}{2021}\natexlab{}.
\newblock \showarticletitle{Quantum computing opportunities in renewable energy}.
\newblock \bibinfo{journal}{\emph{SN Computer Science}} \bibinfo{volume}{2}, \bibinfo{number}{5} (\bibinfo{year}{2021}), \bibinfo{pages}{393}.
\newblock


\bibitem[He et~al\mbox{.}(2016)]%
        {ai-resnet}
\bibfield{author}{\bibinfo{person}{Kaiming He}, \bibinfo{person}{Xiangyu Zhang}, \bibinfo{person}{Shaoqing Ren}, {and} \bibinfo{person}{Jian Sun}.} \bibinfo{year}{2016}\natexlab{}.
\newblock \showarticletitle{Deep residual learning for image recognition}. In \bibinfo{booktitle}{\emph{Proceedings of the IEEE conference on computer vision and pattern recognition}}. \bibinfo{pages}{770--778}.
\newblock


\bibitem[Head-Gordon and Pople(1988)]%
        {HGP}
\bibfield{author}{\bibinfo{person}{Martin Head-Gordon} {and} \bibinfo{person}{John~A Pople}.} \bibinfo{year}{1988}\natexlab{}.
\newblock \showarticletitle{A method for two-electron Gaussian integral and integral derivative evaluation using recurrence relations}.
\newblock \bibinfo{journal}{\emph{The Journal of chemical physics}} \bibinfo{volume}{89}, \bibinfo{number}{9} (\bibinfo{year}{1988}), \bibinfo{pages}{5777--5786}.
\newblock


\bibitem[Helgaker et~al\mbox{.}(2008a)]%
        {QC-review-2}
\bibfield{author}{\bibinfo{person}{Trygve Helgaker}, \bibinfo{person}{Wim Klopper}, {and} \bibinfo{person}{David~P Tew}.} \bibinfo{year}{2008}\natexlab{a}.
\newblock \showarticletitle{Quantitative quantum chemistry}.
\newblock \bibinfo{journal}{\emph{Molecular Physics}} \bibinfo{volume}{106}, \bibinfo{number}{16-18} (\bibinfo{year}{2008}), \bibinfo{pages}{2107--2143}.
\newblock


\bibitem[Helgaker et~al\mbox{.}(2008b)]%
        {helgaker2008quantitative}
\bibfield{author}{\bibinfo{person}{Trygve Helgaker}, \bibinfo{person}{Wim Klopper}, {and} \bibinfo{person}{David~P Tew}.} \bibinfo{year}{2008}\natexlab{b}.
\newblock \showarticletitle{Quantitative quantum chemistry}.
\newblock \bibinfo{journal}{\emph{Molecular Physics}} \bibinfo{volume}{106}, \bibinfo{number}{16-18} (\bibinfo{year}{2008}), \bibinfo{pages}{2107--2143}.
\newblock


\bibitem[Häser and Ahlrichs(1989)]%
        {CPU-Algo-4}
\bibfield{author}{\bibinfo{person}{Marco Häser} {and} \bibinfo{person}{Reinhart Ahlrichs}.} \bibinfo{year}{1989}\natexlab{}.
\newblock \showarticletitle{Improvements on the direct SCF method}.
\newblock \bibinfo{journal}{\emph{Journal of Computational Chemistry}} \bibinfo{volume}{10}, \bibinfo{number}{1} (\bibinfo{year}{1989}), \bibinfo{pages}{104--111}.
\newblock
\urldef\tempurl%
\url{https://doi.org/10.1002/jcc.540100111}
\showDOI{\tempurl}
\showeprint{https://onlinelibrary.wiley.com/doi/pdf/10.1002/jcc.540100111}


\bibitem[Johnson(1975)]%
        {QC-review-3}
\bibfield{author}{\bibinfo{person}{Keith~H Johnson}.} \bibinfo{year}{1975}\natexlab{}.
\newblock \showarticletitle{Quantum chemistry}.
\newblock \bibinfo{journal}{\emph{Annual review of physical chemistry}} \bibinfo{volume}{26}, \bibinfo{number}{1} (\bibinfo{year}{1975}), \bibinfo{pages}{39--57}.
\newblock


\bibitem[Jumper et~al\mbox{.}(2021)]%
        {ai-alphafold}
\bibfield{author}{\bibinfo{person}{John Jumper}, \bibinfo{person}{Richard Evans}, \bibinfo{person}{Alexander Pritzel}, \bibinfo{person}{Tim Green}, \bibinfo{person}{Michael Figurnov}, \bibinfo{person}{Olaf Ronneberger}, \bibinfo{person}{Kathryn Tunyasuvunakool}, \bibinfo{person}{Russ Bates}, \bibinfo{person}{Augustin {\v{Z}}{\'\i}dek}, \bibinfo{person}{Anna Potapenko}, {et~al\mbox{.}}} \bibinfo{year}{2021}\natexlab{}.
\newblock \showarticletitle{Highly accurate protein structure prediction with AlphaFold}.
\newblock \bibinfo{journal}{\emph{Nature}} \bibinfo{volume}{596}, \bibinfo{number}{7873} (\bibinfo{year}{2021}), \bibinfo{pages}{583--589}.
\newblock


\bibitem[Krizhevsky et~al\mbox{.}(2012)]%
        {ai-conv}
\bibfield{author}{\bibinfo{person}{Alex Krizhevsky}, \bibinfo{person}{Ilya Sutskever}, {and} \bibinfo{person}{Geoffrey~E Hinton}.} \bibinfo{year}{2012}\natexlab{}.
\newblock \showarticletitle{Imagenet classification with deep convolutional neural networks}.
\newblock \bibinfo{journal}{\emph{Advances in neural information processing systems}}  \bibinfo{volume}{25} (\bibinfo{year}{2012}).
\newblock


\bibitem[K{\"u}hne et~al\mbox{.}(2020)]%
        {related-work-cp2k}
\bibfield{author}{\bibinfo{person}{Thomas~D K{\"u}hne}, \bibinfo{person}{Marcella Iannuzzi}, \bibinfo{person}{Mauro Del~Ben}, \bibinfo{person}{Vladimir~V Rybkin}, \bibinfo{person}{Patrick Seewald}, \bibinfo{person}{Frederick Stein}, \bibinfo{person}{Teodoro Laino}, \bibinfo{person}{Rustam~Z Khaliullin}, \bibinfo{person}{Ole Sch{\"u}tt}, \bibinfo{person}{Florian Schiffmann}, {et~al\mbox{.}}} \bibinfo{year}{2020}\natexlab{}.
\newblock \showarticletitle{CP2K: An electronic structure and molecular dynamics software package-Quickstep: Efficient and accurate electronic structure calculations}.
\newblock \bibinfo{journal}{\emph{The Journal of Chemical Physics}} \bibinfo{volume}{152}, \bibinfo{number}{19} (\bibinfo{year}{2020}).
\newblock


\bibitem[Li et~al\mbox{.}(2022)]%
        {cpu-stencil-2}
\bibfield{author}{\bibinfo{person}{Kun Li}, \bibinfo{person}{Liang Yuan}, \bibinfo{person}{Yunquan Zhang}, \bibinfo{person}{Yue Yue}, {and} \bibinfo{person}{Hang Cao}.} \bibinfo{year}{2022}\natexlab{}.
\newblock \showarticletitle{An efficient vectorization scheme for stencil computation}. In \bibinfo{booktitle}{\emph{2022 IEEE International Parallel and Distributed Processing Symposium (IPDPS)}}. IEEE, \bibinfo{pages}{650--660}.
\newblock


\bibitem[Liu et~al\mbox{.}(2014)]%
        {CPU-Load-imbalance-1}
\bibfield{author}{\bibinfo{person}{Xing Liu}, \bibinfo{person}{Aftab Patel}, {and} \bibinfo{person}{Edmond Chow}.} \bibinfo{year}{2014}\natexlab{}.
\newblock \showarticletitle{A New Scalable Parallel Algorithm for Fock Matrix Construction}. In \bibinfo{booktitle}{\emph{2014 IEEE 28th International Parallel and Distributed Processing Symposium}}. \bibinfo{pages}{902--914}.
\newblock
\showISSN{1530-2075}
\urldef\tempurl%
\url{https://doi.org/10.1109/IPDPS.2014.97}
\showDOI{\tempurl}


\bibitem[Liu et~al\mbox{.}(2021)]%
        {sc-gb21}
\bibfield{author}{\bibinfo{person}{Yong~(Alexander) Liu}, \bibinfo{person}{Xin~(Lucy) Liu}, \bibinfo{person}{Fang~(Nancy) Li}, \bibinfo{person}{Haohuan Fu}, \bibinfo{person}{Yuling Yang}, \bibinfo{person}{Jiawei Song}, \bibinfo{person}{Pengpeng Zhao}, \bibinfo{person}{Zhen Wang}, \bibinfo{person}{Dajia Peng}, \bibinfo{person}{Huarong Chen}, \bibinfo{person}{Chu Guo}, \bibinfo{person}{Heliang Huang}, \bibinfo{person}{Wenzhao Wu}, {and} \bibinfo{person}{Dexun Chen}.} \bibinfo{year}{2021}\natexlab{}.
\newblock \showarticletitle{Closing the "Quantum Supremacy" Gap: Achieving Real-Time Simulation of a Random Quantum Circuit Using a New Sunway Supercomputer}. In \bibinfo{booktitle}{\emph{Proceedings of the International Conference for High Performance Computing, Networking, Storage and Analysis}} (St. Louis, Missouri) \emph{(\bibinfo{series}{SC '21})}. \bibinfo{publisher}{Association for Computing Machinery}, \bibinfo{address}{New York, NY, USA}, Article \bibinfo{articleno}{3}, \bibinfo{numpages}{12}~pages.
\newblock
\showISBNx{9781450384421}
\urldef\tempurl%
\url{https://doi.org/10.1145/3458817.3487399}
\showDOI{\tempurl}


\bibitem[Manathunga et~al\mbox{.}(2023)]%
        {GPU-QUICK-4}
\bibfield{author}{\bibinfo{person}{Madushanka Manathunga}, \bibinfo{person}{Hasan~Metin Aktulga}, \bibinfo{person}{Andreas~W. Götz}, {and} \bibinfo{person}{Kenneth M.~Jr. Merz}.} \bibinfo{year}{2023}\natexlab{}.
\newblock \showarticletitle{Quantum Mechanics/Molecular Mechanics Simulations on NVIDIA and AMD Graphics Processing Units}.
\newblock \bibinfo{journal}{\emph{Journal of Chemical Information and Modeling}} \bibinfo{volume}{63}, \bibinfo{number}{3} (\bibinfo{year}{2023}), \bibinfo{pages}{711--717}.
\newblock
\urldef\tempurl%
\url{https://doi.org/10.1021/acs.jcim.2c01505}
\showDOI{\tempurl}
\showeprint{https://doi.org/10.1021/acs.jcim.2c01505}
\newblock
\shownote{PMID: 36720086}.


\bibitem[Manathunga et~al\mbox{.}(2021)]%
        {GPU-QUICK-3}
\bibfield{author}{\bibinfo{person}{Madushanka Manathunga}, \bibinfo{person}{Chi Jin}, \bibinfo{person}{Vinícius Wilian~D. Cruzeiro}, \bibinfo{person}{Yipu Miao}, \bibinfo{person}{Dawei Mu}, \bibinfo{person}{Kamesh Arumugam}, \bibinfo{person}{Kristopher Keipert}, \bibinfo{person}{Hasan~Metin Aktulga}, \bibinfo{person}{Kenneth M.~Jr. Merz}, {and} \bibinfo{person}{Andreas~W. Götz}.} \bibinfo{year}{2021}\natexlab{}.
\newblock \showarticletitle{Harnessing the Power of Multi-GPU Acceleration into the Quantum Interaction Computational Kernel Program}.
\newblock \bibinfo{journal}{\emph{Journal of Chemical Theory and Computation}} \bibinfo{volume}{17}, \bibinfo{number}{7} (\bibinfo{year}{2021}), \bibinfo{pages}{3955--3966}.
\newblock
\urldef\tempurl%
\url{https://doi.org/10.1021/acs.jctc.1c00145}
\showDOI{\tempurl}
\showeprint{https://doi.org/10.1021/acs.jctc.1c00145}
\newblock
\shownote{PMID: 34062061}.


\bibitem[Miao and Merz(2013)]%
        {GPU-QUICK-1}
\bibfield{author}{\bibinfo{person}{Yipu Miao} {and} \bibinfo{person}{Kenneth M.~Jr. Merz}.} \bibinfo{year}{2013}\natexlab{}.
\newblock \showarticletitle{Acceleration of Electron Repulsion Integral Evaluation on Graphics Processing Units via Use of Recurrence Relations}.
\newblock \bibinfo{journal}{\emph{Journal of Chemical Theory and Computation}} \bibinfo{volume}{9}, \bibinfo{number}{2} (\bibinfo{year}{2013}), \bibinfo{pages}{965--976}.
\newblock
\urldef\tempurl%
\url{https://doi.org/10.1021/ct300754n}
\showDOI{\tempurl}
\showeprint{https://doi.org/10.1021/ct300754n}
\newblock
\shownote{PMID: 26588740}.


\bibitem[Miao and Merz(2015)]%
        {GPU-QUICK-2}
\bibfield{author}{\bibinfo{person}{Yipu Miao} {and} \bibinfo{person}{Kenneth M.~Jr. Merz}.} \bibinfo{year}{2015}\natexlab{}.
\newblock \showarticletitle{Acceleration of High Angular Momentum Electron Repulsion Integrals and Integral Derivatives on Graphics Processing Units}.
\newblock \bibinfo{journal}{\emph{Journal of Chemical Theory and Computation}} \bibinfo{volume}{11}, \bibinfo{number}{4} (\bibinfo{year}{2015}), \bibinfo{pages}{1449--1462}.
\newblock
\urldef\tempurl%
\url{https://doi.org/10.1021/ct500984t}
\showDOI{\tempurl}
\showeprint{https://doi.org/10.1021/ct500984t}
\newblock
\shownote{PMID: 26574356}.


\bibitem[Mironov et~al\mbox{.}(2019a)]%
        {CPU-GAMESS}
\bibfield{author}{\bibinfo{person}{Vladimir Mironov}, \bibinfo{person}{Alexander Moskovsky}, \bibinfo{person}{Michael D’Mello}, {and} \bibinfo{person}{Yuri Alexeev}.} \bibinfo{year}{2019}\natexlab{a}.
\newblock \showarticletitle{An efficient MPI/OpenMP parallelization of the Hartree–Fock–Roothaan method for the first generation of Intel® Xeon Phi™ processor architecture}.
\newblock \bibinfo{journal}{\emph{The International Journal of High Performance Computing Applications}} \bibinfo{volume}{33}, \bibinfo{number}{1} (\bibinfo{year}{2019}), \bibinfo{pages}{212--224}.
\newblock
\urldef\tempurl%
\url{https://doi.org/10.1177/1094342017732628}
\showDOI{\tempurl}
\showeprint{https://doi.org/10.1177/1094342017732628}


\bibitem[Mironov et~al\mbox{.}(2019b)]%
        {related-work-gamess}
\bibfield{author}{\bibinfo{person}{Vladimir Mironov}, \bibinfo{person}{Alexander Moskovsky}, \bibinfo{person}{Michael D’Mello}, {and} \bibinfo{person}{Yuri Alexeev}.} \bibinfo{year}{2019}\natexlab{b}.
\newblock \showarticletitle{An efficient MPI/OpenMP parallelization of the Hartree–Fock–Roothaan method for the first generation of Intel® Xeon Phi™ processor architecture}.
\newblock \bibinfo{journal}{\emph{The International Journal of High Performance Computing Applications}} \bibinfo{volume}{33}, \bibinfo{number}{1} (\bibinfo{year}{2019}), \bibinfo{pages}{212--224}.
\newblock
\urldef\tempurl%
\url{https://doi.org/10.1177/1094342017732628}
\showDOI{\tempurl}
\showeprint{https://doi.org/10.1177/1094342017732628}


\bibitem[Mohr et~al\mbox{.}(2017)]%
        {cpu-earthquake}
\bibfield{author}{\bibinfo{person}{B Mohr}, \bibinfo{person}{P Raghavan}, \bibinfo{person}{H Fu}, {et~al\mbox{.}}} \bibinfo{year}{2017}\natexlab{}.
\newblock \showarticletitle{18.9-Pflops nonlinear earthquake simulation on Sunway TaihuLight}. In \bibinfo{booktitle}{\emph{Proceedings of the International Conference for High Performance Computing, Networking, Storage and Analysis}}. \bibinfo{pages}{1--12}.
\newblock


\bibitem[Parr and Yang(1995)]%
        {DFT}
\bibfield{author}{\bibinfo{person}{Robert~G Parr} {and} \bibinfo{person}{Weitao Yang}.} \bibinfo{year}{1995}\natexlab{}.
\newblock \showarticletitle{Density-functional theory of the electronic structure of molecules}.
\newblock \bibinfo{journal}{\emph{Annual review of physical chemistry}} \bibinfo{volume}{46}, \bibinfo{number}{1} (\bibinfo{year}{1995}), \bibinfo{pages}{701--728}.
\newblock


\bibitem[Pople and Hehre(1978)]%
        {pople1978computation}
\bibfield{author}{\bibinfo{person}{John~A Pople} {and} \bibinfo{person}{Warren~J Hehre}.} \bibinfo{year}{1978}\natexlab{}.
\newblock \showarticletitle{Computation of electron repulsion integrals involving contracted Gaussian basis functions}.
\newblock \bibinfo{journal}{\emph{J. Comput. Phys.}} \bibinfo{volume}{27}, \bibinfo{number}{2} (\bibinfo{year}{1978}), \bibinfo{pages}{161--168}.
\newblock


\bibitem[Pulay(1980)]%
        {CPU-Algo-2}
\bibfield{author}{\bibinfo{person}{Péter Pulay}.} \bibinfo{year}{1980}\natexlab{}.
\newblock \showarticletitle{Convergence acceleration of iterative sequences. the case of scf iteration}.
\newblock \bibinfo{journal}{\emph{Chemical Physics Letters}} \bibinfo{volume}{73}, \bibinfo{number}{2} (\bibinfo{year}{1980}), \bibinfo{pages}{393--398}.
\newblock
\showISSN{0009-2614}
\urldef\tempurl%
\url{https://doi.org/10.1016/0009-2614(80)80396-4}
\showDOI{\tempurl}


\bibitem[Schi{\o}tz et~al\mbox{.}(1998)]%
        {QC-material-1}
\bibfield{author}{\bibinfo{person}{Jakob Schi{\o}tz}, \bibinfo{person}{Francesco~D Di~Tolla}, {and} \bibinfo{person}{Karsten~W Jacobsen}.} \bibinfo{year}{1998}\natexlab{}.
\newblock \showarticletitle{Softening of nanocrystalline metals at very small grain sizes}.
\newblock \bibinfo{journal}{\emph{Nature}} \bibinfo{volume}{391}, \bibinfo{number}{6667} (\bibinfo{year}{1998}), \bibinfo{pages}{561--563}.
\newblock


\bibitem[Schi{\o}tz and Jacobsen(2003)]%
        {QC-material-2}
\bibfield{author}{\bibinfo{person}{Jakob Schi{\o}tz} {and} \bibinfo{person}{Karsten~W Jacobsen}.} \bibinfo{year}{2003}\natexlab{}.
\newblock \showarticletitle{A maximum in the strength of nanocrystalline copper}.
\newblock \bibinfo{journal}{\emph{Science}} \bibinfo{volume}{301}, \bibinfo{number}{5638} (\bibinfo{year}{2003}), \bibinfo{pages}{1357--1359}.
\newblock


\bibitem[Shan et~al\mbox{.}(2014)]%
        {CPU-Load-imbalance-4}
\bibfield{author}{\bibinfo{person}{Hongzhang Shan}, \bibinfo{person}{Brian Austin}, \bibinfo{person}{Wibe De~Jong}, \bibinfo{person}{Leonid Oliker}, \bibinfo{person}{N.~J. Wright}, {and} \bibinfo{person}{Edoardo Apra}.} \bibinfo{year}{2014}\natexlab{}.
\newblock \showarticletitle{Performance Tuning of Fock Matrix and Two-Electron Integral Calculations for NWChem on Leading HPC Platforms}. In \bibinfo{booktitle}{\emph{High Performance Computing Systems. Performance Modeling, Benchmarking and Simulation}}, \bibfield{editor}{\bibinfo{person}{Stephen~A. Jarvis}, \bibinfo{person}{Steven~A. Wright}, {and} \bibinfo{person}{Simon~D. Hammond}} (Eds.). \bibinfo{publisher}{Springer International Publishing}, \bibinfo{address}{Cham}, \bibinfo{pages}{261--280}.
\newblock
\showISBNx{978-3-319-10214-6}


\bibitem[Shan et~al\mbox{.}(2015)]%
        {CPU-Load-imbalance-5}
\bibfield{author}{\bibinfo{person}{Hongzhang Shan}, \bibinfo{person}{Samuel Williams}, \bibinfo{person}{Wibe de Jong}, {and} \bibinfo{person}{Leonid Oliker}.} \bibinfo{year}{2015}\natexlab{}.
\newblock \showarticletitle{Thread-Level Parallelization and Optimization of NWChem for the Intel MIC Architecture}. In \bibinfo{booktitle}{\emph{Proceedings of the Sixth International Workshop on Programming Models and Applications for Multicores and Manycores}} (San Francisco, California) \emph{(\bibinfo{series}{PMAM '15})}. \bibinfo{publisher}{Association for Computing Machinery}, \bibinfo{address}{New York, NY, USA}, \bibinfo{pages}{58–67}.
\newblock
\showISBNx{9781450334044}
\urldef\tempurl%
\url{https://doi.org/10.1145/2712386.2712391}
\showDOI{\tempurl}


\bibitem[Sun(2015)]%
        {CPU-PySCF-1}
\bibfield{author}{\bibinfo{person}{Qiming Sun}.} \bibinfo{year}{2015}\natexlab{}.
\newblock \showarticletitle{Libcint: An efficient general integral library for Gaussian basis functions}.
\newblock \bibinfo{journal}{\emph{Journal of Computational Chemistry}} \bibinfo{volume}{36}, \bibinfo{number}{22} (\bibinfo{year}{2015}), \bibinfo{pages}{1664--1671}.
\newblock
\urldef\tempurl%
\url{https://doi.org/10.1002/jcc.23981}
\showDOI{\tempurl}
\showeprint{https://onlinelibrary.wiley.com/doi/pdf/10.1002/jcc.23981}


\bibitem[Sun et~al\mbox{.}(2018)]%
        {CPU-PySCF-2}
\bibfield{author}{\bibinfo{person}{Qiming Sun}, \bibinfo{person}{Timothy~C. Berkelbach}, \bibinfo{person}{Nick~S. Blunt}, \bibinfo{person}{George~H. Booth}, \bibinfo{person}{Sheng Guo}, \bibinfo{person}{Zhendong Li}, \bibinfo{person}{Junzi Liu}, \bibinfo{person}{James~D. McClain}, \bibinfo{person}{Elvira~R. Sayfutyarova}, \bibinfo{person}{Sandeep Sharma}, \bibinfo{person}{Sebastian Wouters}, {and} \bibinfo{person}{Garnet Kin-Lic Chan}.} \bibinfo{year}{2018}\natexlab{}.
\newblock \showarticletitle{PySCF: the Python-based simulations of chemistry framework}.
\newblock \bibinfo{journal}{\emph{WIREs Computational Molecular Science}} \bibinfo{volume}{8}, \bibinfo{number}{1} (\bibinfo{year}{2018}), \bibinfo{pages}{e1340}.
\newblock
\urldef\tempurl%
\url{https://doi.org/10.1002/wcms.1340}
\showDOI{\tempurl}
\showeprint{https://wires.onlinelibrary.wiley.com/doi/pdf/10.1002/wcms.1340}


\bibitem[Sun et~al\mbox{.}(2020)]%
        {CPU-PySCF-3}
\bibfield{author}{\bibinfo{person}{Qiming Sun}, \bibinfo{person}{Xing Zhang}, \bibinfo{person}{Samragni Banerjee}, \bibinfo{person}{Peng Bao}, \bibinfo{person}{Marc Barbry}, \bibinfo{person}{Nick~S. Blunt}, \bibinfo{person}{Nikolay~A. Bogdanov}, \bibinfo{person}{George~H. Booth}, \bibinfo{person}{Jia Chen}, \bibinfo{person}{Zhi-Hao Cui}, \bibinfo{person}{Janus~J. Eriksen}, \bibinfo{person}{Yang Gao}, \bibinfo{person}{Sheng Guo}, \bibinfo{person}{Jan Hermann}, \bibinfo{person}{Matthew~R. Hermes}, \bibinfo{person}{Kevin Koh}, \bibinfo{person}{Peter Koval}, \bibinfo{person}{Susi Lehtola}, \bibinfo{person}{Zhendong Li}, \bibinfo{person}{Junzi Liu}, \bibinfo{person}{Narbe Mardirossian}, \bibinfo{person}{James~D. McClain}, \bibinfo{person}{Mario Motta}, \bibinfo{person}{Bastien Mussard}, \bibinfo{person}{Hung~Q. Pham}, \bibinfo{person}{Artem Pulkin}, \bibinfo{person}{Wirawan Purwanto}, \bibinfo{person}{Paul~J. Robinson}, \bibinfo{person}{Enrico Ronca}, \bibinfo{person}{Elvira~R. Sayfutyarova},
  \bibinfo{person}{Maximilian Scheurer}, \bibinfo{person}{Henry~F. Schurkus}, \bibinfo{person}{James E.~T. Smith}, \bibinfo{person}{Chong Sun}, \bibinfo{person}{Shi-Ning Sun}, \bibinfo{person}{Shiv Upadhyay}, \bibinfo{person}{Lucas~K. Wagner}, \bibinfo{person}{Xiao Wang}, \bibinfo{person}{Alec White}, \bibinfo{person}{James~Daniel Whitfield}, \bibinfo{person}{Mark~J. Williamson}, \bibinfo{person}{Sebastian Wouters}, \bibinfo{person}{Jun Yang}, \bibinfo{person}{Jason~M. Yu}, \bibinfo{person}{Tianyu Zhu}, \bibinfo{person}{Timothy~C. Berkelbach}, \bibinfo{person}{Sandeep Sharma}, \bibinfo{person}{Alexander~Yu. Sokolov}, {and} \bibinfo{person}{Garnet Kin-Lic Chan}.} \bibinfo{year}{2020}\natexlab{}.
\newblock \showarticletitle{Recent developments in the PySCF program package}.
\newblock \bibinfo{journal}{\emph{The Journal of Chemical Physics}} \bibinfo{volume}{153}, \bibinfo{number}{2} (\bibinfo{date}{07} \bibinfo{year}{2020}), \bibinfo{pages}{024109}.
\newblock
\showISSN{0021-9606}
\urldef\tempurl%
\url{https://doi.org/10.1063/5.0006074}
\showDOI{\tempurl}
\showeprint{https://pubs.aip.org/aip/jcp/article-pdf/doi/10.1063/5.0006074/16722275/024109\_1\_online.pdf}


\bibitem[Touvron et~al\mbox{.}(2023)]%
        {ai-llama}
\bibfield{author}{\bibinfo{person}{Hugo Touvron}, \bibinfo{person}{Thibaut Lavril}, \bibinfo{person}{Gautier Izacard}, \bibinfo{person}{Xavier Martinet}, \bibinfo{person}{Marie-Anne Lachaux}, \bibinfo{person}{Timoth{\'e}e Lacroix}, \bibinfo{person}{Baptiste Rozi{\`e}re}, \bibinfo{person}{Naman Goyal}, \bibinfo{person}{Eric Hambro}, \bibinfo{person}{Faisal Azhar}, {et~al\mbox{.}}} \bibinfo{year}{2023}\natexlab{}.
\newblock \showarticletitle{Llama: Open and efficient foundation language models}.
\newblock \bibinfo{journal}{\emph{arXiv preprint arXiv:2302.13971}} (\bibinfo{year}{2023}).
\newblock


\bibitem[Ufimtsev and Martinez(2008)]%
        {ufimtsev2008quantum}
\bibfield{author}{\bibinfo{person}{Ivan~S Ufimtsev} {and} \bibinfo{person}{Todd~J Martinez}.} \bibinfo{year}{2008}\natexlab{}.
\newblock \showarticletitle{Quantum chemistry on graphical processing units. 1. Strategies for two-electron integral evaluation}.
\newblock \bibinfo{journal}{\emph{Journal of Chemical Theory and Computation}} \bibinfo{volume}{4}, \bibinfo{number}{2} (\bibinfo{year}{2008}), \bibinfo{pages}{222--231}.
\newblock


\bibitem[Ufimtsev and Martinez(2009a)]%
        {GPU-Ufimtsev-3}
\bibfield{author}{\bibinfo{person}{Ivan~S. Ufimtsev} {and} \bibinfo{person}{Todd~J. Martinez}.} \bibinfo{year}{2009}\natexlab{a}.
\newblock \showarticletitle{Quantum Chemistry on Graphical Processing Units. 2. Direct Self-Consistent-Field Implementation}.
\newblock \bibinfo{journal}{\emph{Journal of Chemical Theory and Computation}} \bibinfo{volume}{5}, \bibinfo{number}{4} (\bibinfo{year}{2009}), \bibinfo{pages}{1004--1015}.
\newblock
\urldef\tempurl%
\url{https://doi.org/10.1021/ct800526s}
\showDOI{\tempurl}
\showeprint{https://doi.org/10.1021/ct800526s}
\newblock
\shownote{PMID: 26609609}.


\bibitem[Ufimtsev and Martinez(2009b)]%
        {GPU-Ufimtsev-4}
\bibfield{author}{\bibinfo{person}{Ivan~S. Ufimtsev} {and} \bibinfo{person}{Todd~J. Martinez}.} \bibinfo{year}{2009}\natexlab{b}.
\newblock \showarticletitle{Quantum Chemistry on Graphical Processing Units. 3. Analytical Energy Gradients, Geometry Optimization, and First Principles Molecular Dynamics}.
\newblock \bibinfo{journal}{\emph{Journal of Chemical Theory and Computation}} \bibinfo{volume}{5}, \bibinfo{number}{10} (\bibinfo{year}{2009}), \bibinfo{pages}{2619--2628}.
\newblock
\urldef\tempurl%
\url{https://doi.org/10.1021/ct9003004}
\showDOI{\tempurl}
\showeprint{https://doi.org/10.1021/ct9003004}
\newblock
\shownote{PMID: 26631777}.


\bibitem[Ufimtsev and Martínez(2008a)]%
        {GPU-Ufimtsev-1}
\bibfield{author}{\bibinfo{person}{Ivan~S. Ufimtsev} {and} \bibinfo{person}{Todd~J. Martínez}.} \bibinfo{year}{2008}\natexlab{a}.
\newblock \showarticletitle{Graphical Processing Units for Quantum Chemistry}.
\newblock \bibinfo{journal}{\emph{Computing in Science \& Engineering}} \bibinfo{volume}{10}, \bibinfo{number}{6} (\bibinfo{date}{Nov} \bibinfo{year}{2008}), \bibinfo{pages}{26--34}.
\newblock
\showISSN{1558-366X}
\urldef\tempurl%
\url{https://doi.org/10.1109/MCSE.2008.148}
\showDOI{\tempurl}


\bibitem[Ufimtsev and Martínez(2008b)]%
        {GPU-Ufimtsev-2}
\bibfield{author}{\bibinfo{person}{Ivan~S. Ufimtsev} {and} \bibinfo{person}{Todd~J. Martínez}.} \bibinfo{year}{2008}\natexlab{b}.
\newblock \showarticletitle{Quantum Chemistry on Graphical Processing Units. 1. Strategies for Two-Electron Integral Evaluation}.
\newblock \bibinfo{journal}{\emph{Journal of Chemical Theory and Computation}} \bibinfo{volume}{4}, \bibinfo{number}{2} (\bibinfo{year}{2008}), \bibinfo{pages}{222--231}.
\newblock
\urldef\tempurl%
\url{https://doi.org/10.1021/ct700268q}
\showDOI{\tempurl}
\showeprint{https://doi.org/10.1021/ct700268q}
\newblock
\shownote{PMID: 26620654}.


\bibitem[Valeev(2022)]%
        {CPU-Libint}
\bibfield{author}{\bibinfo{person}{E.~F. Valeev}.} \bibinfo{year}{2022}\natexlab{}.
\newblock \bibinfo{title}{Libint: A library for the evaluation of molecular integrals of many-body operators over Gaussian functions}.
\newblock \bibinfo{howpublished}{http://libint.valeyev.net/}.
\newblock
\newblock
\shownote{version 2.8.0}.


\bibitem[Valiev et~al\mbox{.}(2010a)]%
        {CPU-distributed-1}
\bibfield{author}{\bibinfo{person}{M. Valiev}, \bibinfo{person}{E.J. Bylaska}, \bibinfo{person}{N. Govind}, \bibinfo{person}{K. Kowalski}, \bibinfo{person}{T.P. Straatsma}, \bibinfo{person}{H.J.J. {Van Dam}}, \bibinfo{person}{D. Wang}, \bibinfo{person}{J. Nieplocha}, \bibinfo{person}{E. Apra}, \bibinfo{person}{T.L. Windus}, {and} \bibinfo{person}{W.A. {de Jong}}.} \bibinfo{year}{2010}\natexlab{a}.
\newblock \showarticletitle{NWChem: A comprehensive and scalable open-source solution for large scale molecular simulations}.
\newblock \bibinfo{journal}{\emph{Computer Physics Communications}} \bibinfo{volume}{181}, \bibinfo{number}{9} (\bibinfo{year}{2010}), \bibinfo{pages}{1477--1489}.
\newblock
\showISSN{0010-4655}
\urldef\tempurl%
\url{https://doi.org/10.1016/j.cpc.2010.04.018}
\showDOI{\tempurl}


\bibitem[Valiev et~al\mbox{.}(2010b)]%
        {related-work-nwchem}
\bibfield{author}{\bibinfo{person}{M. Valiev}, \bibinfo{person}{E.J. Bylaska}, \bibinfo{person}{N. Govind}, \bibinfo{person}{K. Kowalski}, \bibinfo{person}{T.P. Straatsma}, \bibinfo{person}{H.J.J. {Van Dam}}, \bibinfo{person}{D. Wang}, \bibinfo{person}{J. Nieplocha}, \bibinfo{person}{E. Apra}, \bibinfo{person}{T.L. Windus}, {and} \bibinfo{person}{W.A. {de Jong}}.} \bibinfo{year}{2010}\natexlab{b}.
\newblock \showarticletitle{NWChem: A comprehensive and scalable open-source solution for large scale molecular simulations}.
\newblock \bibinfo{journal}{\emph{Computer Physics Communications}} \bibinfo{volume}{181}, \bibinfo{number}{9} (\bibinfo{year}{2010}), \bibinfo{pages}{1477--1489}.
\newblock
\showISSN{0010-4655}
\urldef\tempurl%
\url{https://doi.org/10.1016/j.cpc.2010.04.018}
\showDOI{\tempurl}


\bibitem[Vaswani et~al\mbox{.}(2017)]%
        {ai-transformer}
\bibfield{author}{\bibinfo{person}{Ashish Vaswani}, \bibinfo{person}{Noam Shazeer}, \bibinfo{person}{Niki Parmar}, \bibinfo{person}{Jakob Uszkoreit}, \bibinfo{person}{Llion Jones}, \bibinfo{person}{Aidan~N Gomez}, \bibinfo{person}{{\L}ukasz Kaiser}, {and} \bibinfo{person}{Illia Polosukhin}.} \bibinfo{year}{2017}\natexlab{}.
\newblock \showarticletitle{Attention is all you need}.
\newblock \bibinfo{journal}{\emph{Advances in neural information processing systems}}  \bibinfo{volume}{30} (\bibinfo{year}{2017}).
\newblock


\bibitem[Wang et~al\mbox{.}(2020)]%
        {cpu-fft}
\bibfield{author}{\bibinfo{person}{Qinglin Wang}, \bibinfo{person}{Dongsheng Li}, \bibinfo{person}{Xiandong Huang}, \bibinfo{person}{Siqi Shen}, \bibinfo{person}{Songzhu Mei}, {and} \bibinfo{person}{Jie Liu}.} \bibinfo{year}{2020}\natexlab{}.
\newblock \showarticletitle{Optimizing FFT-based convolution on ARMv8 multi-core CPUs}. In \bibinfo{booktitle}{\emph{European Conference on Parallel Processing}}. Springer, \bibinfo{pages}{248--262}.
\newblock


\bibitem[Yasuda(2008)]%
        {GPU-Yasuda}
\bibfield{author}{\bibinfo{person}{Koji Yasuda}.} \bibinfo{year}{2008}\natexlab{}.
\newblock \showarticletitle{Two-electron integral evaluation on the graphics processor unit}.
\newblock \bibinfo{journal}{\emph{Journal of Computational Chemistry}} \bibinfo{volume}{29}, \bibinfo{number}{3} (\bibinfo{year}{2008}), \bibinfo{pages}{334--342}.
\newblock
\urldef\tempurl%
\url{https://doi.org/10.1002/jcc.20779}
\showDOI{\tempurl}
\showeprint{https://onlinelibrary.wiley.com/doi/pdf/10.1002/jcc.20779}


\end{thebibliography}

\end{sloppypar}
\end{document}